\documentstyle[12pt,psfig]{article}
\newcommand{\norm}[1]{{\protect\normalsize{#1}}}
\newcommand{\LAP}
{{\small E}\norm{N}{\large S}{\Large L}{\large A}\norm{P}{\small P}}

\newcommand{\vs}[1]{\rule[- #1 mm]{0mm}{#1 mm}}

\newbox\hdbox%
\newcount\hdrows%
\newcount\multispancount%
\newcount\ncase%
\newcount\ncols% This is the number of primary text columns in the table.
\newcount\nrows%
\newcount\nspan%
\newcount\ntemp%
\newdimen\hdsize%
\newdimen\newhdsize%
\newdimen\parasize%
\newdimen\spreadwidth%
\newdimen\thicksize%
\newdimen\thinsize%
\newdimen\tablewidth%
\newif\ifcentertables%
\newif\ifendsize%
\newif\iffirstrow%
\newif\iftableinfo%
\newtoks\dbt%
\newtoks\hdtks%
\newtoks\savetks%
\newtoks\tableLETtokens%
\newtoks\tabletokens%
\newtoks\widthspec%
\immediate\write15{%
CP SMSG GJMSINK TEXTABLE --> TABLE MACROS V. 851121 JOB = \jobname%
}%
\tableinfotrue%
\catcode`\@=11%  Allows use of "@" in macro names, like PLAIN.TEX does.
\def\tstrut{\vrule height3.1ex depth1.2ex width0pt}%
\def\and{\char`\&}%  Allows us to get an `&' in the text.  This is the
\def\tablerule{\noalign{\hrule height\thinsize depth0pt}}%
\def\thickrule{\noalign{\hrule height\thicksize depth0pt}}%
\def\ctr#1{\hfil\ #1\hfil}%
\tablewidth=-\maxdimen%
\spreadwidth=-\maxdimen%
\def\tabskipglue{0pt plus 1fil minus 1fil}%
\centertablestrue%
\gdef\ARGS{########}%  Produces the correct number of #'s in the preamble
\gdef\headerARGS{####}%  Same as \ARGS, but used in \header macros.
\def\@mpersand{&}%  Allows us to get alignment tab characters later
{\catcode`\|=13%  Make |'s locally active.
\gdef\letbarzero{\let|0}%  Globally define a macro that allows us to
\gdef\letbartab{\def|{&&}}%
\gdef\letvbbar{\let\vb|}%
}%  End of locally active |'s.
{\catcode`\&=4%  Make these alignment tabs.
\def\ampskip{&\omit\hfil&}%  This local macro skips a vertical rule.
\catcode`\&=13%  Now make &'s into active macros.
\let&0%  This allows us to expand \ampskip in the next \xdef without
\xdef\letampskip{\def&{\ampskip}}%
\gdef\letnovbamp{\let\novb&\let\tab&}
}%  End of locally active &'s.
\def\begintable{%  Here we make |'s and &'s active characters so we can
   \begingroup%
   \catcode`\|=13\letbartab\letvbbar%
   \catcode`\&=13\letampskip\letnovbamp%
   \def\multispan##1{%  We must redefine \multispan to count the number
      \omit \mscount##1%
      \multiply\mscount\tw@\advance\mscount\m@ne%
      \loop\ifnum\mscount>\@ne \sp@n\repeat%
   }%  End of \multispan macro.
   \def\|{%
      &\omit\widevline&%
   }%
   \ruledtable%  Now we call \ruledtable to do the real work.
}%  End of \begintable macro.
\long\def\ruledtable#1\endtable{%
   \offinterlineskip%  Needed to make rules touch each other.
   \tabskip 0pt%  Needed for same reason as \offinterlineskip.
   \def\widevline{\vrule width\thicksize}%  Make outer \vrule's wider.
   \def\endrow{\@mpersand\omit\hfil\crnorm\@mpersand}%
   \def\crthick{\@mpersand\crnorm\thickrule\@mpersand}%
   \def\crthickneg##1{\@mpersand\crnorm\thickrule
          \noalign{{\skip0=##1\vskip-\skip0}}\@mpersand}%
   \def\crnorule{\@mpersand\crnorm\@mpersand}%
   \def\crnoruleneg##1{\@mpersand\crnorm
          \noalign{{\skip0=##1\vskip-\skip0}}\@mpersand}%
   \let\nr=\crnorule%  A shorter abbreviation.
   \def\endtable{\@mpersand\crnorm\thickrule}%
   \let\crnorm=\cr%  Allows us to use \cr for our own purposes.
   \edef\cr{\@mpersand\crnorm\tablerule\@mpersand}%
   \def\crneg##1{\@mpersand\crnorm\tablerule
          \noalign{{\skip0=##1\vskip-\skip0}}\@mpersand}%
   \let\ctneg=\crthickneg
   \let\nrneg=\crnoruleneg
   \the\tableLETtokens%  Get the user's extra \let's, if any.
   \tabletokens={&#1}%  We add an extra alignment tab to the beginning
   \countROWS\tabletokens\into\nrows%
   \countCOLS\tabletokens\into\ncols%
   \advance\ncols by -1%
   \divide\ncols by 2%
   \advance\nrows by 1%
   \iftableinfo %
      \immediate\write16{[Nrows=\the\nrows, Ncols=\the\ncols]}%
   \fi%
   \ifcentertables
      \ifhmode \par\fi%  Make sure we are in vertical mode.
      \hbox to \hsize{% The final table comes out as an \hbox of width the
      \hss%  The final table will be centered left-to-right.
   \else %
      \hbox{%
   \fi
      \vbox{%
         \makePREAMBLE{\the\ncols}%  Generate the preamble.
         \edef\next{\preamble}%  This line and the next line force the
         \let\preamble=\next%    expansion of all \ARGS tokens into the
         \makeTABLE{\preamble}{\tabletokens}%  Go do the \halign here.
      }%  End of \vbox.
      \ifcentertables \hss}\else }\fi%  Finish the centering effect.
   \endgroup%  Return all local macros and parameters to their outside
   \tablewidth=-\maxdimen%  Reset \tablewidth to normal.
   \spreadwidth=-\maxdimen% Same for \spreadwidth.
}%  End of macro \ruledtable.
\def\makeTABLE#1#2{%  Does an \halign for the \ruledtable macro.
   {%  Start of local parameter values.
   \let\ifmath0%     These macros would cause trouble if they were to be
   \let\header0%     expanded in the following \xdef; we \let them be
   \let\multispan0%  equal to a digit, because digits can't be expanded.
   \ncase=0%
   \ifdim\tablewidth>-\maxdimen \ncase=1\fi%
   \ifdim\spreadwidth>-\maxdimen \ncase=2\fi%
   \relax%  This \relax is absolutely necessary, without it the following
   \ifcase\ncase %
      \widthspec={}%
   \or %
      \widthspec=\expandafter{\expandafter t\expandafter o%
                 \the\tablewidth}%
   \else %
      \widthspec=\expandafter{\expandafter s\expandafter p\expandafter r%
                 \expandafter e\expandafter a\expandafter d%
                 \the\spreadwidth}%
   \fi %
   \xdef\next{%  We must force the preamble to be expanded BEFORE the
      \halign\the\widthspec{%
      #1%  This is the preamble text.
      \noalign{\hrule height\thicksize depth0pt}%  Makes the top \hrule.
      \the#2\endtable%  This is the main body.
      }%  End of \halign.
   }%  End of \next.
   }%  End of local values.
   \next%  This \next must be outside of the local values, because now
}%  End of macro \makeTABLE.
\def\makePREAMBLE#1{%  This macro generates the necessary preamble for a
   \ncols=#1%  Get the number of columns desired.
   \begingroup%  Start local parameter definitions.
   \let\ARGS=0%  This is the key to the whole thing; it prevents \ARGS
   \edef\xtp{\widevline\ARGS\tabskip\tabskipglue%
   &\ctr{\ARGS}\tstrut}%  A 1-column preamble.  Gets the sizing right.
   \advance\ncols by -1%  One column has been generated; decrement the
   \loop%  Append as many further columns as needed to the preamble.
      \ifnum\ncols>0 %
      \advance\ncols by -1%
      \edef\xtp{\xtp&\vrule width\thinsize\ARGS&\ctr{\ARGS}}%
   \repeat
   \xdef\preamble{\xtp&\widevline\ARGS\tabskip0pt%
   \crnorm}%  Adds the last \vrule.
   \endgroup%  End of local parameters.
}%  End of macro \makePREAMBLE.
\def\countROWS#1\into#2{%  This counts the number of rows in #1 by
   \let\countREGISTER=#2%
   \countREGISTER=0%
   \expandafter\ROWcount\the#1\endcount%
}%
\def\ROWcount{%
   \afterassignment\subROWcount\let\next= %
}%
\def\subROWcount{%
   \ifx\next\endcount %
      \let\next=\relax%
   \else%
      \ncase=0%
      \ifx\next\cr %
         \global\advance\countREGISTER by 1%
         \ncase=0%
      \fi%
      \ifx\next\endrow %
         \global\advance\countREGISTER by 1%
         \ncase=0%
      \fi%
      \ifx\next\crthick %
         \global\advance\countREGISTER by 1%
         \ncase=0%
      \fi%
      \ifx\next\crnorule %
         \global\advance\countREGISTER by 1%
         \ncase=0%
      \fi%
      \ifx\next\crthickneg %
         \global\advance\countREGISTER by 1%
         \ncase=0%
      \fi%
      \ifx\next\crnoruleneg %
         \global\advance\countREGISTER by 1%
         \ncase=0%
      \fi%
      \ifx\next\crneg %
         \global\advance\countREGISTER by 1%
         \ncase=0%
      \fi%
      \ifx\next\header %
         \ncase=1%
      \fi%
      \relax%
      \ifcase\ncase %
         \let\next\ROWcount%
      \or %
         \let\next\argROWskip%
      \else %
      \fi%
   \fi%
   \next%
}%  End of macro \subROWcount.
\def\counthdROWS#1\into#2{%
\dvr{10}%
   \let\countREGISTER=#2%
   \countREGISTER=0%
\dvr{11}%
\dvr{13}%
   \expandafter\hdROWcount\the#1\endcount%
\dvr{12}%
}%
\def\hdROWcount{%
   \afterassignment\subhdROWcount\let\next= %
}%
\def\subhdROWcount{%
   \ifx\next\endcount %
      \let\next=\relax%
   \else%
      \ncase=0%
      \ifx\next\cr %
         \global\advance\countREGISTER by 1%
         \ncase=0%
      \fi%
      \ifx\next\endrow %
         \global\advance\countREGISTER by 1%
         \ncase=0%
      \fi%
      \ifx\next\crthick %
         \global\advance\countREGISTER by 1%
         \ncase=0%
      \fi%
      \ifx\next\crnorule %
         \global\advance\countREGISTER by 1%
         \ncase=0%
      \fi%
      \ifx\next\header %
         \ncase=1%
      \fi%
\relax%
      \ifcase\ncase %
         \let\next\hdROWcount%
      \or%
         \let\next\arghdROWskip%
      \else %
      \fi%
   \fi%
   \next%
}%
{\catcode`\|=13\letbartab
\gdef\countCOLS#1\into#2{%
   \let\countREGISTER=#2%
   \global\countREGISTER=0%
   \global\multispancount=0%
   \global\firstrowtrue
   \expandafter\COLcount\the#1\endcount%
   \global\advance\countREGISTER by 3%
   \global\advance\countREGISTER by -\multispancount
}%
\gdef\COLcount{%
   \afterassignment\subCOLcount\let\next= %
}%
{\catcode`\&=13%
\gdef\subCOLcount{%
   \ifx\next\endcount %
      \let\next=\relax%
   \else%
      \ncase=0%
      \iffirstrow
         \ifx\next& %
            \global\advance\countREGISTER by 2%
            \ncase=0%
         \fi%
         \ifx\next\span %
            \global\advance\countREGISTER by 1%
            \ncase=0%
         \fi%
         \ifx\next| %
            \global\advance\countREGISTER by 2%
            \ncase=0%
         \fi
         \ifx\next\|
            \global\advance\countREGISTER by 2%
            \ncase=0%
         \fi
         \ifx\next\multispan
            \ncase=1%
            \global\advance\multispancount by 1%
         \fi
         \ifx\next\header
            \ncase=2%
         \fi
         \ifx\next\cr       \global\firstrowfalse \fi
         \ifx\next\endrow   \global\firstrowfalse \fi
         \ifx\next\crthick  \global\firstrowfalse \fi
         \ifx\next\crnorule \global\firstrowfalse \fi
         \ifx\next\crnoruleneg \global\firstrowfalse \fi
         \ifx\next\crthickneg  \global\firstrowfalse \fi
         \ifx\next\crneg       \global\firstrowfalse \fi
      \fi%  End of \iffirstrow.
\relax%\out{subCOL-->  ncase=[\the\ncase]}
      \ifcase\ncase %
         \let\next\COLcount%
      \or %
         \let\next\spancount%
      \or %
         \let\next\argCOLskip%
      \else %
      \fi %
   \fi%
   \next%
}%
\gdef\argROWskip#1{%
   \let\next\ROWcount \next%
}%  End of macro \argskip.
\gdef\arghdROWskip#1{%
   \let\next\ROWcount \next%
}%  End of macro \arghdROWskip.
\gdef\argCOLskip#1{%
   \let\next\COLcount \next%
}%  End of macro \argskip.
}%  End of active &'s.
}%  End of active |'s.
\def\spancount#1{%\out{spancount--->\meaning#1}
   \nspan=#1\multiply\nspan by 2\advance\nspan by -1%
   \global\advance \countREGISTER by \nspan
   \let\next\COLcount \next}%
\def\dvr#1{\relax}%
\def\header#1{%
\dvr{1}{\let\cr=\@mpersand%
\hdtks={#1}%
\counthdROWS\hdtks\into\hdrows%
\advance\hdrows by 1%
\ifnum\hdrows=0 \hdrows=1 \fi%
\dvr{5}\makehdPREAMBLE{\the\hdrows}%
\dvr{6}\getHDdimen{#1}%
{\parindent=0pt\hsize=\hdsize{\let\ifmath0%
\xdef\next{\valign{\headerpreamble #1\crnorm}}}\dvr{7}\next\dvr{8}%
}%
}\dvr{2}}%  End of macro \header.
\def\makehdPREAMBLE#1{%This macro generates the necessary preamble for a
\dvr{3}%
\hdrows=#1%  Get the number of columns desired.
{%  Start local parameter definitions.
\let\headerARGS=0%
\let\cr=\crnorm%
\edef\xtp{\vfil\hfil\hbox{\headerARGS}\hfil\vfil}%
\advance\hdrows by -1%  One row has been generated; decrement the
\loop%  Append as many further rows as needed to the preamble.
\ifnum\hdrows>0%
\advance\hdrows by -1%
\edef\xtp{\xtp&\vfil\hfil\hbox{\headerARGS}\hfil\vfil}%
\repeat%
\xdef\headerpreamble{\xtp\crcr}%
}%  End of local parameters.
\dvr{4}}%  End of \makehdPREAMBLE.
\def\getHDdimen#1{%
\hdsize=0pt%
\getsize#1\cr\end\cr%
}%  End of macro getHDdimen.
\def\getsize#1\cr{%
\endsizefalse\savetks={#1}%
\expandafter\lookend\the\savetks\cr%
\relax \ifendsize \let\next\relax \else%
\setbox\hdbox=\hbox{#1}\newhdsize=1.0\wd\hdbox%
\ifdim\newhdsize>\hdsize \hdsize=\newhdsize \fi%
\let\next\getsize \fi%
\next%
}%
\def\lookend{\afterassignment\sublookend\let\looknext= }%
\def\sublookend{\relax%
\ifx\looknext\cr %
\let\looknext\relax \else %
   \relax
   \ifx\looknext\end \global\endsizetrue \fi%
   \let\looknext=\lookend%
    \fi \looknext%
}%
\def\tablelet#1{%
   \tableLETtokens=\expandafter{\the\tableLETtokens #1}%
}%
\catcode`\@=12%  Change @'s back to their normal category code.
\begin{document}
\renewcommand{\thefootnote}{\fnsymbol{footnote}}
\newpage
\pagestyle{empty}
\setcounter{page}{0}

%%%%%%%%%%%%%%%%%%%%%%%%%%%%%%%%%%
%%%%%%%%%%%  ENTETE ENSLAPP  %%%%%%%%%%%%
%%%%%%%%%%%%%%%%%%%%%%%%%%%%%%%%%%

%\input BoxedEPS
%\SetOzTeXEPSFSpecial
%\HideDisplacementBoxes
\null
\begin{minipage}{4.9cm}
\begin{center}
{\bf
G{\sc\bf roupe} d'A{\sc\bf nnecy}\\
\ \\
Laboratoire d'Annecy-le-Vieux de Physique des Particules}
\end{center}
\end{minipage}
\hfill
%\BoxedEPSF{enslapp.epsf}
\hfill
\begin{minipage}{4.2cm}
\begin{center}
{\bf
G{\sc\bf roupe} de L{\sc\bf yon}\\
\ \\
Ecole Normale Sup\'erieure de Lyon}
\end{center}
\end{minipage}

\begin{center}
\rule{14cm}{.42mm}
\end{center}
%%%%%%%%%%%%%%%%%%%%%%%%%%%%%%%%%%
%%%%%%%%%%%  ENTETE ENSLAPP FIN %%%%%%%%%%
%%%%%%%%%%%%%%%%%%%%%%%%%%%%%%%%%%

\begin{center}

{\Large {\bf HADRONIC PHOTON--PHOTON}}\\[0,5cm]
{\Large {\bf INTERACTIONS  AT HIGH ENERGIES}}

\vs{5}

{\bf R.Engel}\\
{\em Institut f\"ur Theoretische Physik, Universit\"at Leipzig, D--04109
Leipzig
Germany \\
and Fachbereich Physik, Universit\"at Siegen, D--57068 Siegen, Germany}

\vs{0,5}

{\em and}

\vs{0,5}

{\bf J.Ranft}\\

{\em Laboratoire de Physique Th\'eorique \LAP\footnote{URA 14-36
du CNRS, associ\'ee \`a l'Ecole Normale Sup\'erieure de Lyon et \`a
l'Universit\'e de Savoie.} \\Groupe d'Annecy: LAPP, Chemin de Bellevue, BP
110, \\F-74941
Annecy-le-Vieux Cedex, France.}

\end{center}
\vs{5}

\centerline{ {\bf Abstract}}

\indent

Photon-photon collisions are investigated in the framework of the
two-component
Dual Parton Model. The model contains contributions from direct,
resolved soft and resolved hard interactions. All free
parameters of the model are determined in fits to hadron-hadron
and photon-hadron cross section data. The model is shown to
agree well to hadron production data from hadron-hadron and
photon-hadron collisions. The multiparticle production in
hadron-hadron, photon-hadron and photon-photon collisions as
predicted by the model is compared. Strong differences are only
found as function of the transverse momentum variable. The
hadron production in photon-photon collisions at present and
future electron-positron colliders is studied using photon
spectra according to the equivalent photon approximation,
according to beamstrahlung and according to backscattered laser
radiation.

\vfill
\rightline{\LAP-A-540/95}
\rightline{Siegen SI 95-08}
\rightline{September 1995}

\newpage
\pagestyle{plain}
\renewcommand{\thefootnote}{\arabic{footnote}}

\section{Introduction}

The photon, in it's high-energy interactions with hadrons,
behaves very much like a hadron, however with cross sections
reduced strongly against pure hadronic cross sections. Besides
this soft hadronic interaction, usually described using the Vector
Dominance Model (VDM), the photon has a direct (QED) pointlike
interaction with the hadronic constituents and it has the
resolved hard interaction between it's hadronic constituents and
the hadronic constituents of the target. At moderate energies
these hard interactions of the photons do not change significantly
the general picture of photon-hadron and photon-photon
interactions, this is often forgotten if only the hard part of
the photon interaction is discussed.
Even at high energies, hadronic interactions of photons are
characterized by soft multiparticle production.
Since the soft component of hadron
production cannot be understood purely on the basis of
perturbative QCD, one has to rely on models to calculate the
multiparticle final states.
The Dual Parton Model (DPM) (a recent review is given in
Ref.~\cite{capdpm})
has been very successfully describing soft hadronic processes in
hadron-hadron collisions.
Observations
like rapidity plateaus and average transverse momenta
rising with energy,
KNO scaling violation,
transverse momentum-multiplicity correlations and
minijets pointed out
that soft and hard processes are closely related.
These properties were understood within the two-component Dual Parton
Model for hadron-hadron interactions by Aurenche et al.
\cite{DTUJETPR92,CTK87,DTUJETZP91,DTUJET92a,DTUJET92b,DTUJETDIFF,DTUJET93}.

Assuming an
universal behavior of soft hadronic interactions, it is possible to
extent the Dual Parton Model to hadronic interactions involving photons.
First studies of photon-hadron interactions in the framework of
the two-component Dual Parton Model were done by Engel
\cite{engelgpsig,Engel94c,engelgp}. This reaction was studied within
other models by various authors, one example is the work by
Schuler and Sj\"ostrand \cite{Schuler93b,schusjo94}.

Here we apply the model described in
\cite{engelgpsig,engelgp} to the study of hadronic
photon-photon interactions.
In Section II we give a brief introduction to the  Dual
Parton Model used to describe photon-hadron and
photon-photon collisions, a complete account of the model can be found
in \cite{engelgpsig,engelgp}. An overview of the model realization in
the Monte Carlo event generator {\sc phojet} is given.
We study with {\sc phojet} hadron-hadron and photon-hadron
collisions and compare to data in Section III.
In Section IV  we compare the properties of minimum bias
hadron production as calculated with {\sc phojet} in
hadron-hadron, photon-hadron and photon-photon collisions.
In Section V we discuss the properties of hadron production via
photon-photon collisions at
present and future electron-positron colliders.
A Summary is given in Section VI.

\section{The event generator {\sc phojet}}

The realization of the Dual Parton Model with a hard and a soft
component in {\sc phojet} is similar to the event generator {\sc dtujet}-93
\cite{DTUJETPR92,DTUJET93}
simulating $p$-$p$ and $p$-$\bar p$ collisions up to very high
energies.

In the following, we restrict our discussion of the model
to the basic ideas rather
than giving the complete expressions for all the formulae and quantities
entering the model. More detailed descriptions of the model are given
in \cite{engelgpsig,engelgp}.

In the model, the dual nature of the photon is taken into account
by considering the physical photon state as a
superposition of a "bare photon" and virtual hadronic states having the
same quantum numbers $J^{PC} = 1^{--}$ as the photon.
Since the properties of the high-mass hadronic fluctuations
of the photon are not well known, it is necessary to introduce some
approximations for calculations. To keep the model as simple as possible,
two generic hadronic states $| q\bar q\rangle$ and
$| {q\bar q}^\star\rangle$ have been introduced to describe the hadronic
piece of the photon. The low-mass state $| q\bar q\rangle$ corresponds
to the superposition of the vector mesons $\rho$, $\omega$ and $\phi$
and a $\pi^+\pi^-$ background. The state $| {q\bar q}^\star\rangle$ is
used as an approximation for hadronic states with higher masses (e.g.\
$\rho^\prime$, $\omega^\prime$ or $\rho^{\prime\prime}$).
The physical photon reads
\begin{equation}
|\gamma \rangle = \sqrt{Z_3}\ |\gamma_{\mbox{\scriptsize bare}} \rangle +
|\gamma_{\mbox{\scriptsize had}} \rangle
\end{equation}
with
\begin{equation}
Z_3 = 1 - \frac{e^2}{f_{q\bar{q}}^2} -
\frac{e^2}{f_{q\bar{q}^\star}^2} \hspace{0.8cm}\mbox{and}\hspace{0.8cm}
|\gamma_{\mbox{\scriptsize had}} \rangle =
\frac{e}{f_{q\bar{q}}} \ |q\bar{q} \rangle +
\frac{e}{f_{q\bar{q}^\star}} \ |q\bar{q}^\star \rangle
\end{equation}
where $e$ denotes the elementary charge.

The interactions of the hadronic fluctuations are described
within the Dual Parton Model in terms of
reggeon (${{I\!\!R}}$) and pomeron (${{I\!\!P}}$) exchanges.
For soft processes, photon-hadron duality is used.
The energy-dependence of the reggeon and
pomeron amplitude is assumed to be the same for all hadronic processes.
Therefore, data on hadron-hadron and photon-hadron cross sections
can be used to determine the parameters necessary to describe soft
photon-photon interactions.
However, one does not expect that this photon-hadron universality holds
for processes involving
short distances (high transverse momenta). There, long- and short-living
fluctuations can contribute.

In order to use the framework of the QCD-improved parton model with
lowest order matrix elements \cite{CKR77,Duke82a},
the pomeron exchange is artificially
subdivided into processes involving only {\it soft}
processes and all the other processes with at least one large momentum
transfer ({\it hard} processes).
In the model, soft and hard processes are distinguished by applying
a transverse momentum cutoff $p_\perp^{\mbox{\scriptsize cutoff}}$ to
the partons.
On Born-graph level, for example, the photon-photon cross sections read:\\
(i) reggeon and pomeron exchange (soft processes only)
\begin{eqnarray}
\sigma^{\mbox{\scriptsize tot}}_{{I\!\!P}}(s) =
\left(\frac{e^2}{f^2_{q\bar q}}\right)^2\;
g_{{{I\!\!P}},q\bar{q}}(0)\;g_{{{I\!\!P}},q\bar{q}}(0)\;
\left(\frac{s}{s_0}\right)^{\Delta_{{I\!\!P}}}
\nonumber\\
\sigma^{\mbox{\scriptsize tot}}_{{I\!\!R}}(s) =
\left(\frac{e^2}{f^2_{q\bar q}}\right)^2\;
g_{{{I\!\!R}},q\bar{q}}(0)\;g_{{{I\!\!R}},q\bar{q}}(0)\;
\left(\frac{s}{s_0}\right)^{\Delta_{{I\!\!R}}}
\label{pom-cs}
\end{eqnarray}
with $\Delta_{{I\!\!P}} = \alpha_{{I\!\!P}}(0) - 1$ and
$\Delta_{{I\!\!R}} = \alpha_{{I\!\!R}}(0) - 1$.
Here we denote with $\alpha_{{I\!\!P}}(0)$ ($\alpha_{{I\!\!R}}(0)$)
the pomeron (reggeon)
intercept, and with $g_{{{I\!\!P}},q\bar{q}}$ ($g_{{{I\!\!R}},q\bar{q}}$)
the couplings of the pomeron (reggeon) to the hadronic
$q\bar{q}$-fluctuations.\\
(ii) hard resolved photon-photon interaction
\begin{equation}
\sigma^{\mbox{\scriptsize hard}}_{\mbox{\scriptsize res}}(s,p_\perp^{
\mbox{\scriptsize cutoff}}) =
\int dx_1 dx_2
d\hat{t} \sum_{i,j,k,l} \frac{1}{1+\delta_{k,l}} f_{\gamma,i}(x_1,Q^2)
f_{\gamma,j}(x_2,Q^2) \frac{d\sigma_{i,j\rightarrow
k,l}^{\mbox{\scriptsize QCD}}(\hat{s},\hat{t})}{d\hat{t}}
\Theta(p_\perp-p_\perp^{\mbox{\scriptsize cutoff}}),
\label{hard-res}
\end{equation}
where $f_{\gamma,i}(x_1,Q^2)$ is the distribution of the parton
$i$ in the photon and the
sum includes all possible
parton configurations $i,j,k,l$.\\
(iii) single direct interactions
\begin{equation}
\sigma_{\mbox{\scriptsize s-dir}}(s,p_\perp^{\mbox{\scriptsize cutoff}}) =
\int dx d\hat{t} \sum_{i,k,l}
f_{\gamma,i}(x,Q^2) \frac{d\sigma_{\gamma,i\rightarrow
k,l}^{\mbox{\scriptsize QCD}}(\hat{s},\hat{t})}{d\hat{t}}
\Theta(p_\perp-p_\perp^{\mbox{\scriptsize cutoff}})
\label{s-dir}
\end{equation}
(iv) double direct interactions
\begin{equation}
\sigma_{\mbox{\scriptsize d-dir}}(s,p_\perp^{
\mbox{\scriptsize cutoff}}) = \int d\hat{t} \sum_{k}
\frac{d\sigma_{\gamma,\gamma\rightarrow
k,\bar{k}}(\hat{s},\hat{t})}{d\hat{t}}\Theta(p_\perp-p_\perp^{
\mbox{\scriptsize cutoff}}).
\label{d-dir}
\end{equation}
For simplicity, we have written the
cross section formulae only for the low-mass state
$|q\bar{q}\rangle$. Similar expressions are used for
interactions involving the $|q\bar{q}^\star\rangle$ state.
If not explicitly stated, all the calculations have been done using
the leading order GRV parton distribution functions for the
proton \cite{GRV92a} and the photon \cite{GRV92b}.

Assuming
Gaussian distributions in impact parameter space, the
amplitudes for the different processes can be calculated from the cross
sections given above.

The amplitudes corresponding to the one-pomeron exchange between the
hadronic fluctuations are
unitarized applying a two-channel eikonal formalism similar to
\cite{Capella76,DTUJETPR92}.
In impact parameter representation, the eikonalized scattering amplitude
for resolved photon interactions has the structure
\begin{equation}
a_{\mbox{\scriptsize res}}(s,B) = \frac{i}{2}
\left(\frac{e^2}{f^2_{q\bar q}}\right)^2\;
\left( 1 - e^{-\chi(s,B)}\right)
\label{eff-amp}
\end{equation}
with the eikonal function
\begin{equation}
\chi (s,B)=\chi_{S}(s,B)+\chi_{H}(s,B)+\chi_{D}(s,B)+\chi_{C}(s,B).
\end{equation}
Here, $\chi_{i}(s,B)$ denotes the contributions from the different
Born graphs: (S) soft part of the pomeron and reggeon, (H) hard part
of the pomeron
(D) triple- and loop-pomeron, (C) double-pomeron graphs.
To get the photon-photon scattering amplitude, the resolved
and the direct amplitude given by perturbative QCD are summed up.
The complete expressions will be given in \cite{engelgp}.

\begin{figure}
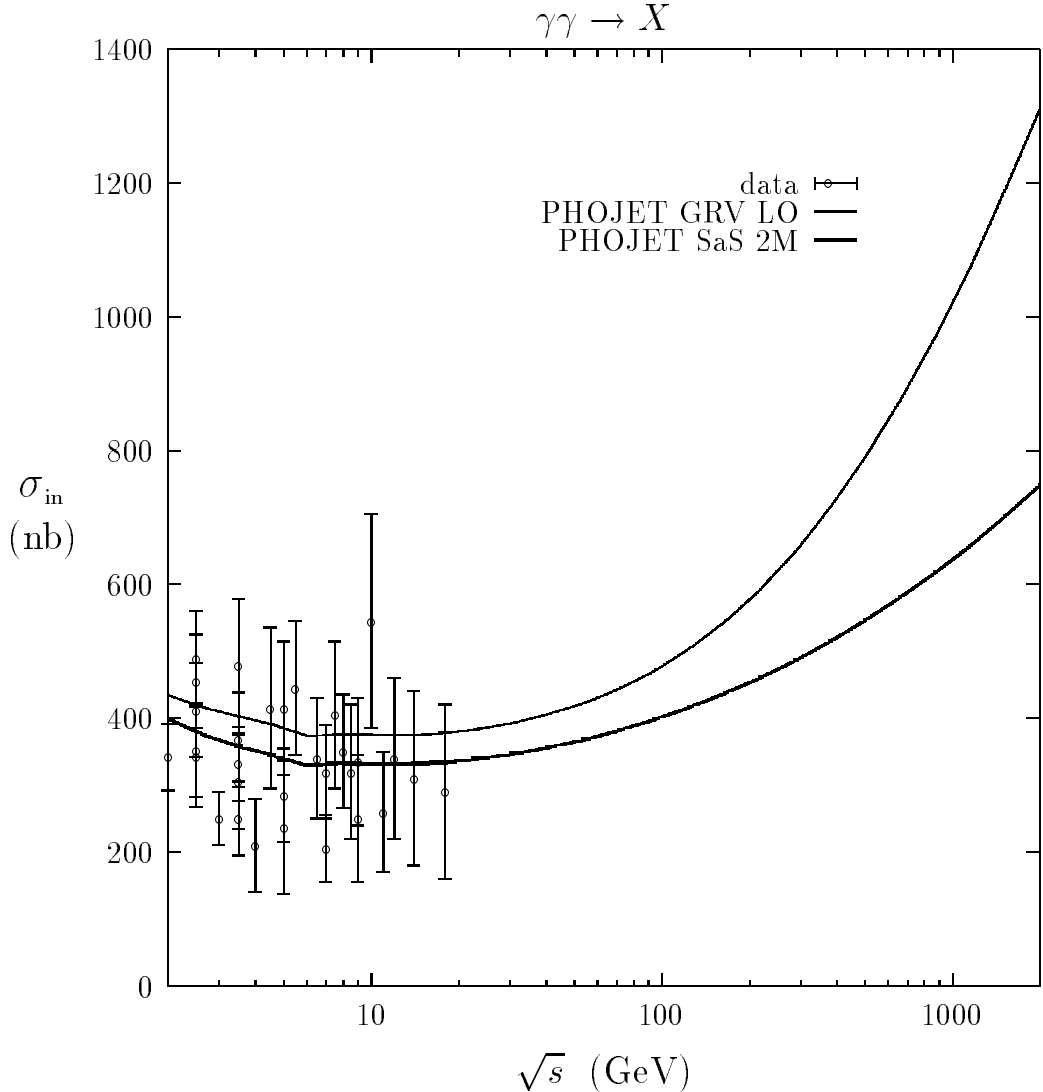

\caption{
Inelastic photon-photon cross sections as calculated in the
model compared with experimental data at low energies
\protect\cite{Feindt86,Aihara90a,Behrend91a,Berger84a,Bintinger85a}.
The two curves from the
model were calculated using the GRV LO photon structure
function \protect\cite{GRV92b} and the SaS 2M photon structure function
\protect\cite{Schuler95a}. The differences between both curves at high
energy demonstrate the uncertainties of the predictions due to the
limited data available on the photon structure function.
Our curve  calculated
with the SaS 2M structure function agrees practically with  the
cross section calculated with the same structure function but
using another model by Schuler and Sj\"ostrand
\protect\cite{schusjo94a}.
\label{ggxtot}
}
\end{figure}

\begin{figure}
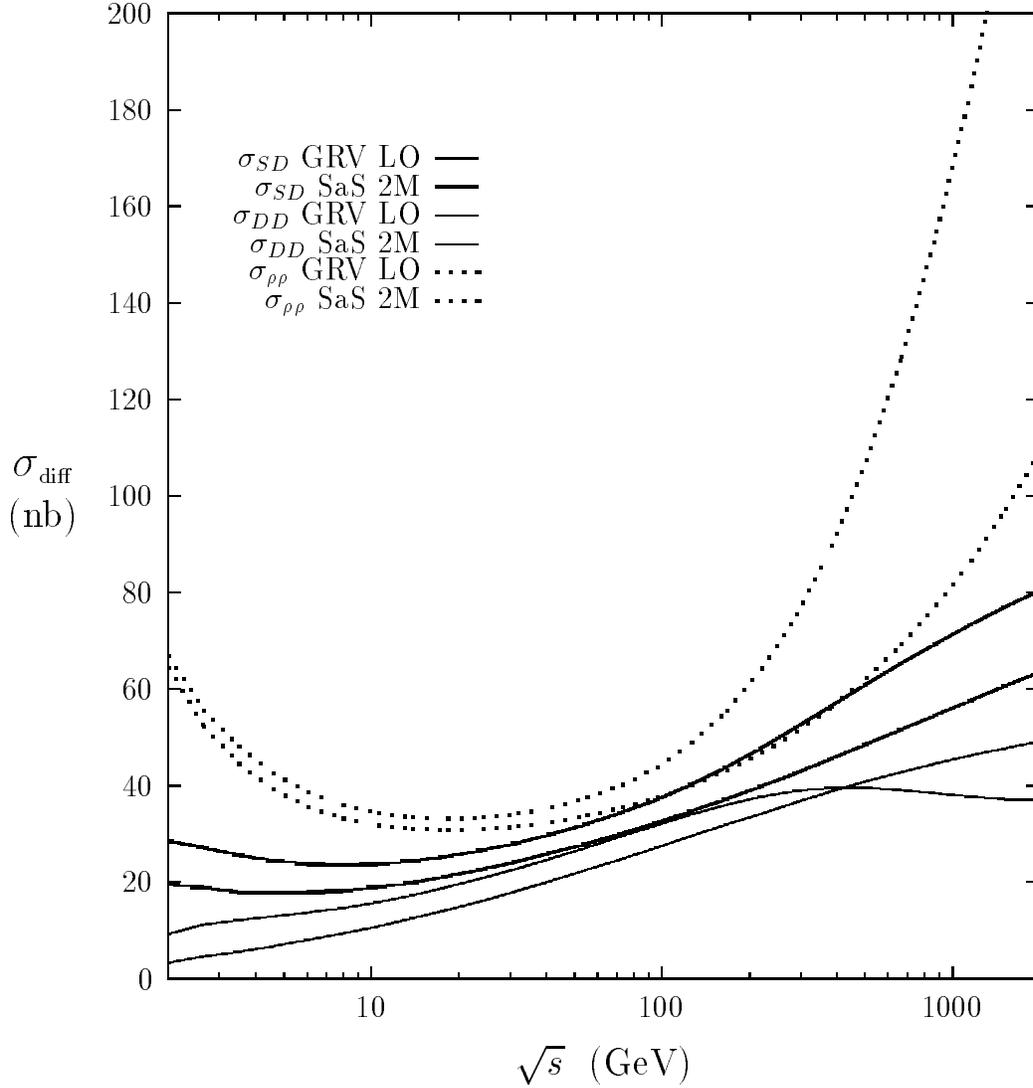


\caption{
Diffractive cross sections as calculated with {\sc phojet}
using the GRV LO photon structure
function \protect\cite{GRV92b} and the SaS 2M photon structure function
\protect\cite{Schuler95a}. The upper curve is for each of the three cross
sections the one obtained with the GRV LO structure function.
\label{ggxdiff}
}
\end{figure}

The probabilities to find a photon in one of the generic hadronic
states, the coupling constants to the reggeon and pomeron, and the
effective reggeon and pomeron intercepts cannot be
determined by basic
principles. These quantities are treated as free parameters.
It was shown in \cite{engelgpsig} that it is possible to fix the free
parameters by a global fit to proton-proton and photon-proton cross
sections and elastic slope parameters. In Fig.~\ref{ggxtot} we show
the model predictions for the inelastic photon-photon cross section
(including quasi-elastic vector meson production). The diffractive
cross sections of quasi-elastic vector meson production ($\rho$, $\omega$
and $\phi$), single diffraction dissociation and double diffraction
dissociation are given in Fig.~\ref{ggxdiff}. In order to show the
strong dependence of the model extrapolations on the parton densities of
the photon, the cross sections have been calculated with two different
parametrizations of the parton distribution functions.

Once the free parameters are determined,
the probabilities for the different final
state configurations are calculated from the discontinuity of the
elastic photon-photon scattering amplitude (optical theorem).
The total discontinuity
can be expressed as a sum of graphs with $k_c$ soft pomeron cuts, $l_c$
hard pomeron cuts, $m_c$ triple- or loop-pomeron cuts, and $n_c$
double-pomeron cuts by applying the Abramovski-Gribov-Kancheli cutting
rules \cite{Abramovski73,TerMartirosyan73}.
In impact parameter space one gets for the inelastic cross
section
\begin{equation}
\sigma (k_{c},l_{c},m_{c},n_{c},s,B)=\frac{(2\chi_{S})^{k_{c}}}{k_{c}!}
\frac{(2\chi_{H})^{l_{c}}}{l_{c}!}\frac{(2\chi_{D})^{m_{c}}}{m_{c}!}
\frac{(2\chi_{C})^{n_{c}}}{n_{c}!}\exp[-2\chi (s,B)]
\label{cutpro}
\end{equation}
with
\begin{equation}
\int d^2B \sum_{k_c+l_c+m_c+n_c=1}^{\infty} \sigma
(k_{c},l_{c},m_{c},n_{c},s,B) \approx \sigma_{\mbox{\scriptsize tot}}
-\sigma_{\mbox{\scriptsize qel}},
\end{equation}
where $\sigma_{\mbox{\scriptsize tot}}$ and
$\sigma_{\mbox{\scriptsize qel}}$ denote the total cross section
and the cross section of quasi-elastic vector meson production, respectively.
We use here the conventions of \cite{engelgpsig} treating the triple- and
loop-pomeron cross sections (and hence $\chi_{D}$) as negative quantities.
In \cite{DTUJETPR92} the negative sign is explicitly written in
the cross section formulae (\ref{cutpro}).
Since the triple-, loop-, and double-pomeron graphs are objects
involving several pomerons, a further resummation is done
\cite{DTUJETPR92,engelgp} to allow for the probability interpretation of
Eq.~(\ref{cutpro}).

In the Monte Carlo realization of the model, the different final state
configurations are sampled from Eq.~(\ref{cutpro}). For pomeron
cuts involving a hard scattering, the complete parton kinematics and
flavors/colors are sampled according to the Parton Model using a method
similar to \cite{Hahn90}, extended to direct processes.
For pomeron cuts without hard large
momentum transfer, the partonic interpretation of the Dual Parton Model
is used: photons or mesons are split into a quark-antiquark pair whereas
baryons are approximated by a quark-diquark pair.
The longitudinal momentum
fractions of the partons
are given by
Regge asymptotics \cite{Capella80a,Capella80b,Kaidalov82a,Kaidalov82b}.
One obtains for the valence quark ($x$) and diquark ($1-x$)
distribution inside the proton
\begin{equation}
\rho(x) \sim \frac{1}{\sqrt{x}} (1-x)^{1.5}
\label{x-baryon}
\end{equation}
and for the quark antiquark distribution inside the photon
\begin{equation}
\rho(x) \sim \frac{1}{\sqrt{x (1-x)}} .
\label{x-meson}
\end{equation}
For multiple interaction events, the sea quark momenta are sampled from a
\begin{equation}
\rho(x) \sim \frac{1}{x}
\end{equation}
distribution. Note that due to energy-momentum conservation,
this distribution is influenced by
the $x$-distributions of the valence partons and asymmetric multiple
interaction effects (for example, in  photon-proton scattering). This
will be discussed in detail elsewhere \cite{Engel95d}.
The transverse
momenta of the soft partons are sampled from the
distribution
\begin{equation}
\frac{d^2 N_s}{d^2 p_\perp} \sim \exp\left( -\beta p_\perp \right).
\end{equation}
The energy-dependent slope parameter $\beta$ is calculated from
the requirement to have a smooth transition between
the transverse momentum distributions of the
soft constituents and the hard scattered partons.

For diffraction dissociation or double-pomeron scattering,
the parton configurations are generated using the
same ideas described above applied to  pomeron-\\
photon/hadron/pomeron scattering processes.
Hence, a diffractive triple-pomeron or loop-pomeron
cut can involve hard scattering subprocesses resulting in a
rapidity gap event with jets.
According to the kinematics of the triple- or loop-pomeron graphs, the
mass of the diffractively dissociating systems is sampled from a
 $1/M_D^{2\alpha_{{I\!\!P}}(0)}$ distribution. The momentum transfer in
diffraction
is obtained from an exponential distribution with mass-dependent slope
(see Ref.~\cite{engelgpsig}).
For the parton distributions of the pomeron, the
CKMT parametrization with a hard gluonic component \cite{Capella95a,Engel95c}
is used.

\begin{figure}
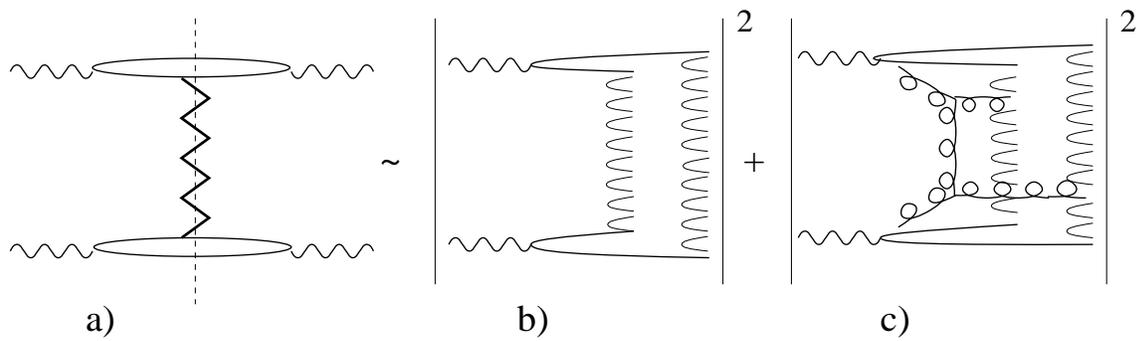

\caption{
Unitarity cut of a one-pomeron graph: the unitarity sum including all
possible final states is subdivided into final states with
low-\protect$p_\perp$ partons and into final states with at least one
parton satisfying \protect$p_\perp \ge p_\perp^{\mbox{\scriptsize cutoff}}$.
\label{ggpom1p}
}
\end{figure}

Finally, the fragmentation of the sampled partonic final states is done
by forming color neutral strings between the partons according to the
color flow.
For soft processes, the color flow is approximated using the expansion of
QCD for large numbers of colors and
flavors. This leads to the two-chain configuration
characterizing a cut pomeron (as shown in Fig.~\ref{ggpom1p} (a),(b))
and a one-chain system for a cut reggeon.
In hard interactions the color flow is taken from the matrix elements
directly \cite{Bengtsson84}.
The leading contributions of the matrix elements give a two-chain structure
which corresponds to a cut pomeron.
For example, a cut of a single hard pomeron
graph (hard gluon-gluon scattering) is shown in Fig.~\ref{ggpom1p} (c).
This method is also applied to the direct photon interactions.

The chains are fragmented using the Lund fragmentation code {\sc jetset} 7.3
 \cite{sj87b}.

\section{Hadron-hadron  and photon-hadron
collisions with {\sc phojet}}

Hadron production in hadron-hadron collisions has been extensively
studied within the two-component Dual Parton Model using the
{\sc dtujet} model
\cite{DTUJETPR92,DTUJET93} in $p$-$p$ and $\bar p$-$p$ collisions
and using the {\sc dpmjet-ii} model \cite{DPMJETII} in
hadron-hadron, hadron-nucleus and nucleus-nucleus collisions.
Hadron production in photon-hadron collisions is being studied
in detail
using the {\sc phojet} model by Engel \cite{Engel94c,engelgp}. We
present here only a few comparisons of {\sc phojet} results with
hadron-hadron and photon-hadron data in order to illustrate,
that the model as formulated in {\sc phojet} is very well able
to describe these channels. This is certainly required, if we
want to apply the model to  photon-photon reactions, where
nearly no
experimental data are available. For photon-photon collisions
we have to rely on the
predictive power of the model.

\subsection{\bf  Hadron-hadron collisions}

\begin{figure}
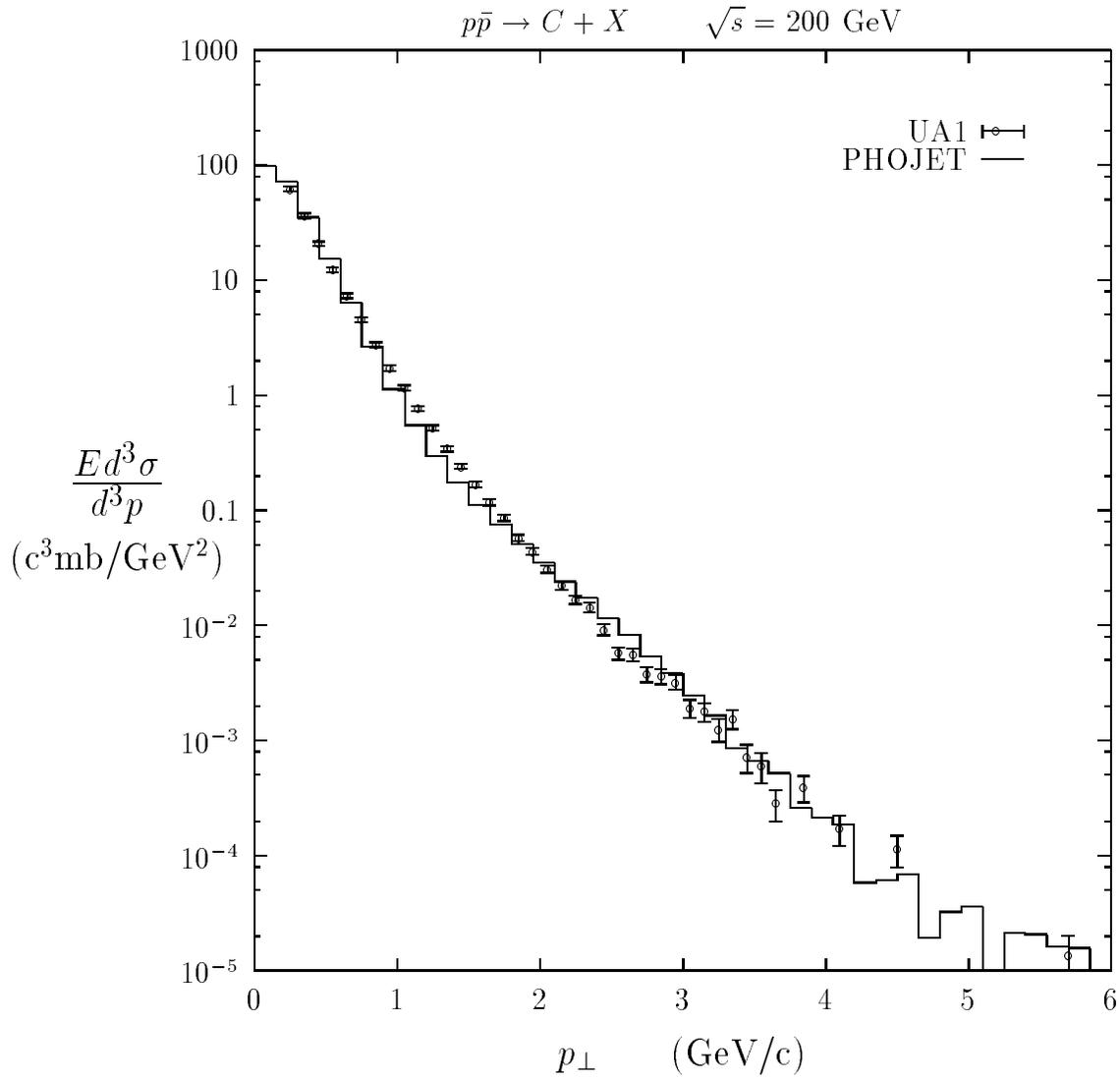

\caption{
Comparison of transverse momentum distributions of charged
hadrons with
collider data at $\protect\sqrt s$ = 200 GeV
\protect\cite{Albajar90}. The calculation uses the Dual Parton Model
code {\sc phojet}.
\label{ua1dndpt}
}
\end{figure}

\begin{figure}
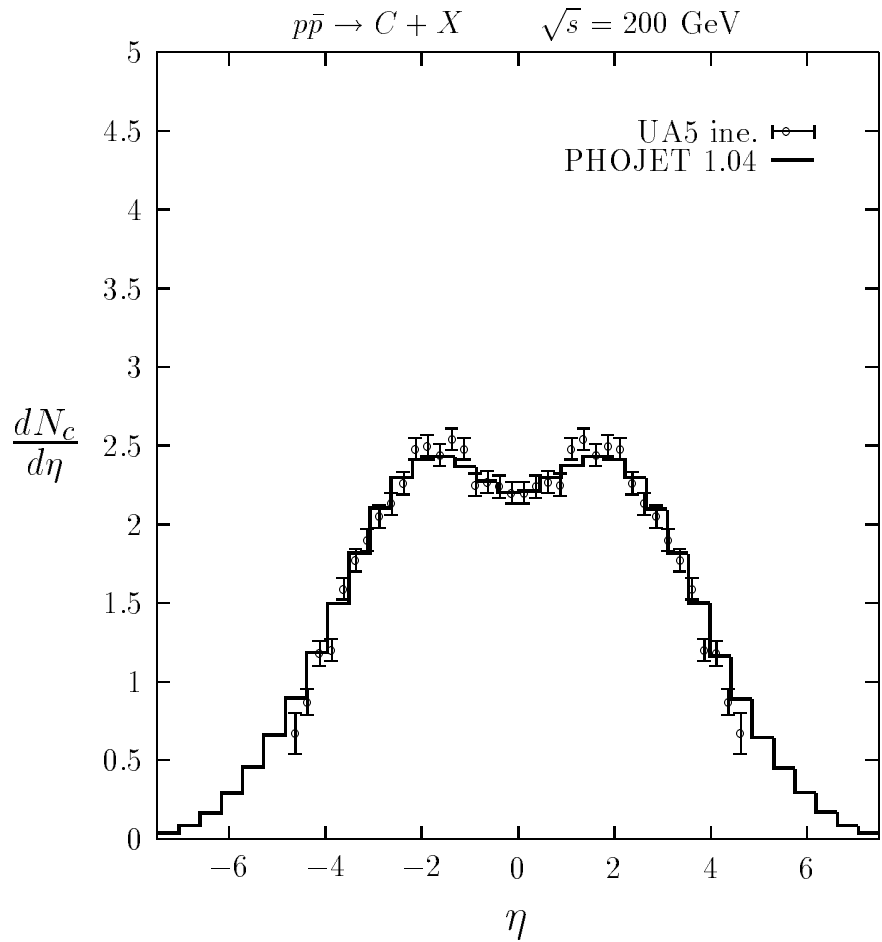

\caption{
 Pseudorapidity distributions  of charged hadrons produced in
 $\bar p$-$p$ collisions as calculated with {\sc phojet}
are  compared to collider data from the UA-5 Collaboration
\protect\cite{Alner86b,Abe90}
for the  energy
 $\protect\sqrt s =$ 200 GeV.
\label{ua5psine}
}
\end{figure}

In Figs. \ref{ua1dndpt} and \ref{ua5psine} we compare the
transverse momentum distributions and pseudorapidity
distributions with the results from collider experiments
\cite{Albajar90,Alner86b,Abe90}. The rise of the plateau with
the collision energy
is understood within the model by one of its most important
ingredients: the production of multiple soft interactions and
multiple hard interactions (minijets) rises with energy.
The transverse momentum distributions as measured at
collider energies show besides the soft, exponentially
damped, low $p_{\perp}$ component of
 hadron production the rise  of the
perturbative hard component with increasing energy.
This becomes more and more
important with rising energy for minimum-bias hadron production.

Discussing the changes in hadron production with rising energy,
we should indicate, that {\sc phojet} in its present version
should not be applied for hadron-hadron, photon-hadron or
photon-photon collisions  at collision energies beyond $\sqrt
s$ = 1 TeV. One possible  way, how to apply a model with minijets
in the TeV energy region of future proton-proton
supercolliders has been implemented in {\sc dtujet}-93
\cite{DTUJET93}. This could also be done in {\sc phojet},
however photon-hadron or photon-photon collisions in the
multi-TeV energy range seem not to be  foreseen for the near future.

\begin{figure}
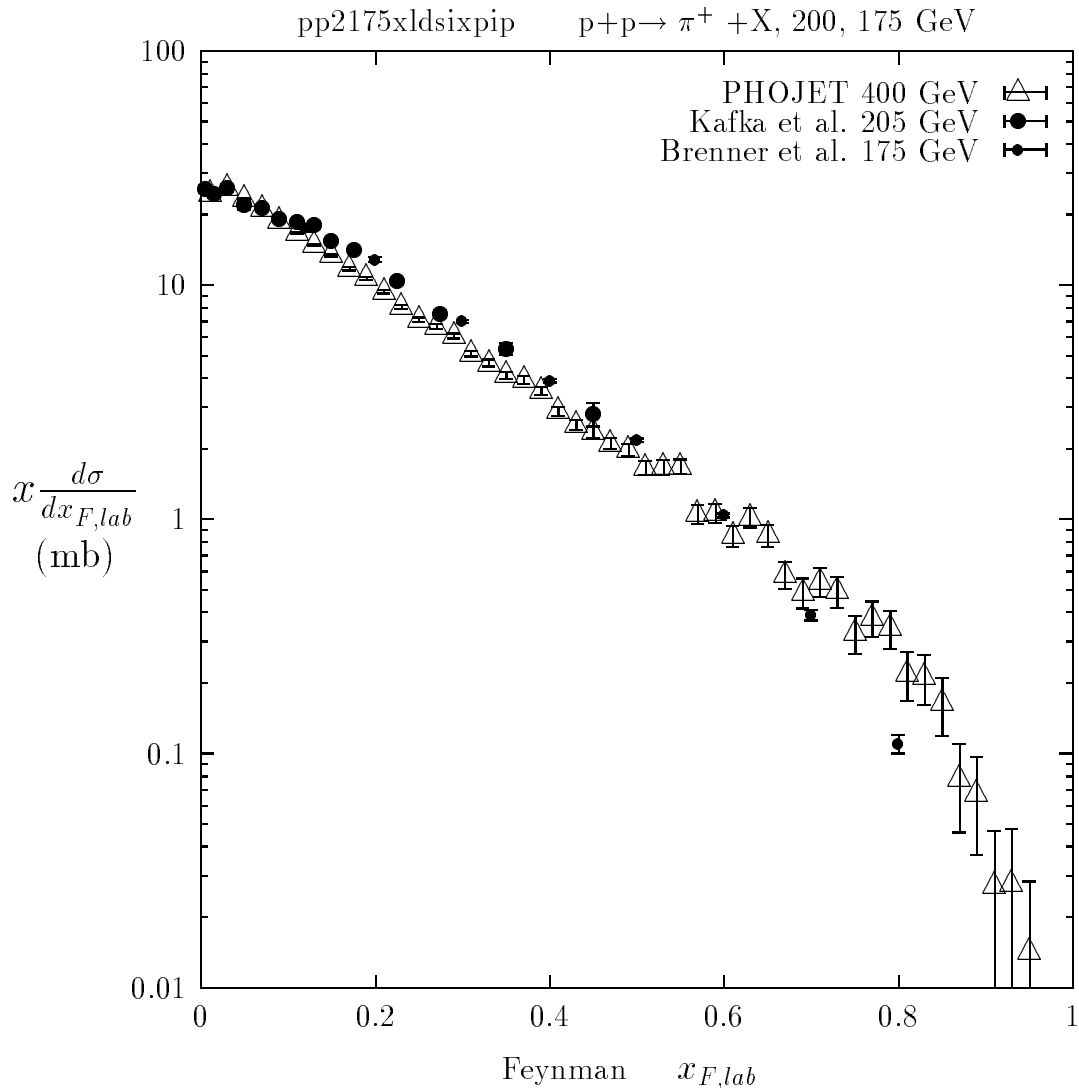

\caption{
Comparison of Feynman $x_F$ distributions of $\pi^+$-mesons
produced in proton-proton collisions at 205  and 175 GeV. The
experimental data are from Kafka  et al. \protect\cite{Kafka}
 and from Brenner et al. \protect\cite{Brenner}.
 The data from both experiments agree rather
well with each other.
 The calculation uses the Dual Parton Model
{\sc phojet}.
\label{pp2175xldsixpip}
}
\end{figure}

\begin{figure}
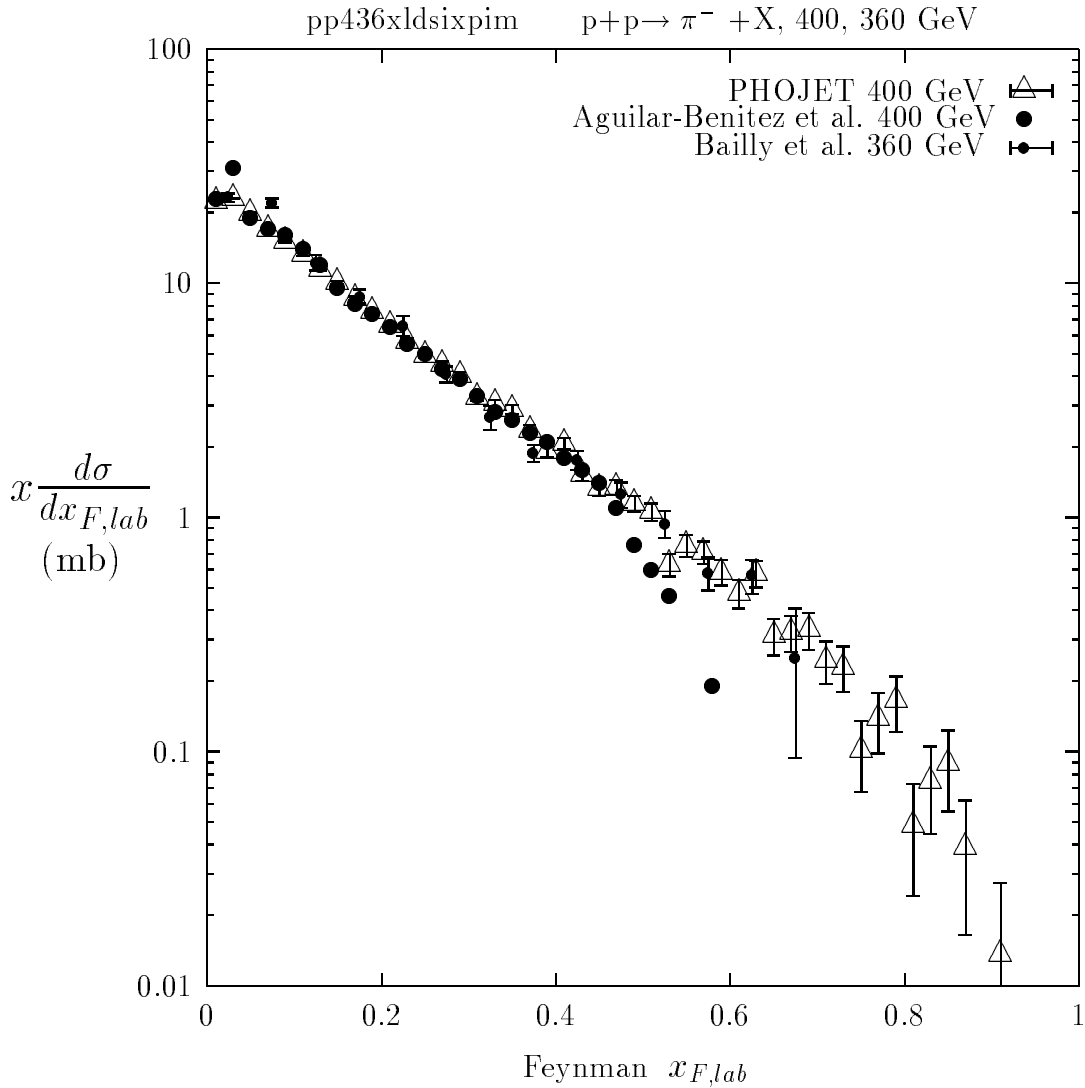

\caption{
Comparison of Feynman $x_F$ distributions of $\pi^-$-mesons
produced in proton-proton collisions at 400  and 360 GeV. The
experimental data are from the Aguilar-Benitez et al.
\protect\cite{pp400exp} and from the EHS-RCBC Collaboration
\protect\cite{Baillypppi}. The data from both experiments agree rather
well with each other, in fact most of the data points of
\protect\cite{Baillypppi} are below the \protect\cite{pp400exp} data.
 The calculation uses the Dual Parton Model
{\sc phojet}.
\label{pp436xldsixpim}
}
\end{figure}

In Figs.
\ref{pp2175xldsixpip} and
\ref{pp436xldsixpim} we compare {\sc phojet} calculations to data
on longitudinal distributions in the fragmentation region using
the Feynman scaling variable $x_F$.
Feynman
$x_F$-distributions have so far only been measured at fixed
target energies, at lab. energies well below 1 TeV.
It has been shown elsewhere
\cite{DPMJETII}, that the Dual Parton Model shows outside the
central region (rise of plateau) and the very forward
fragmentation region (where the diffractive component is
dominating) a very good Feynman scaling behavior.

\begin{figure}
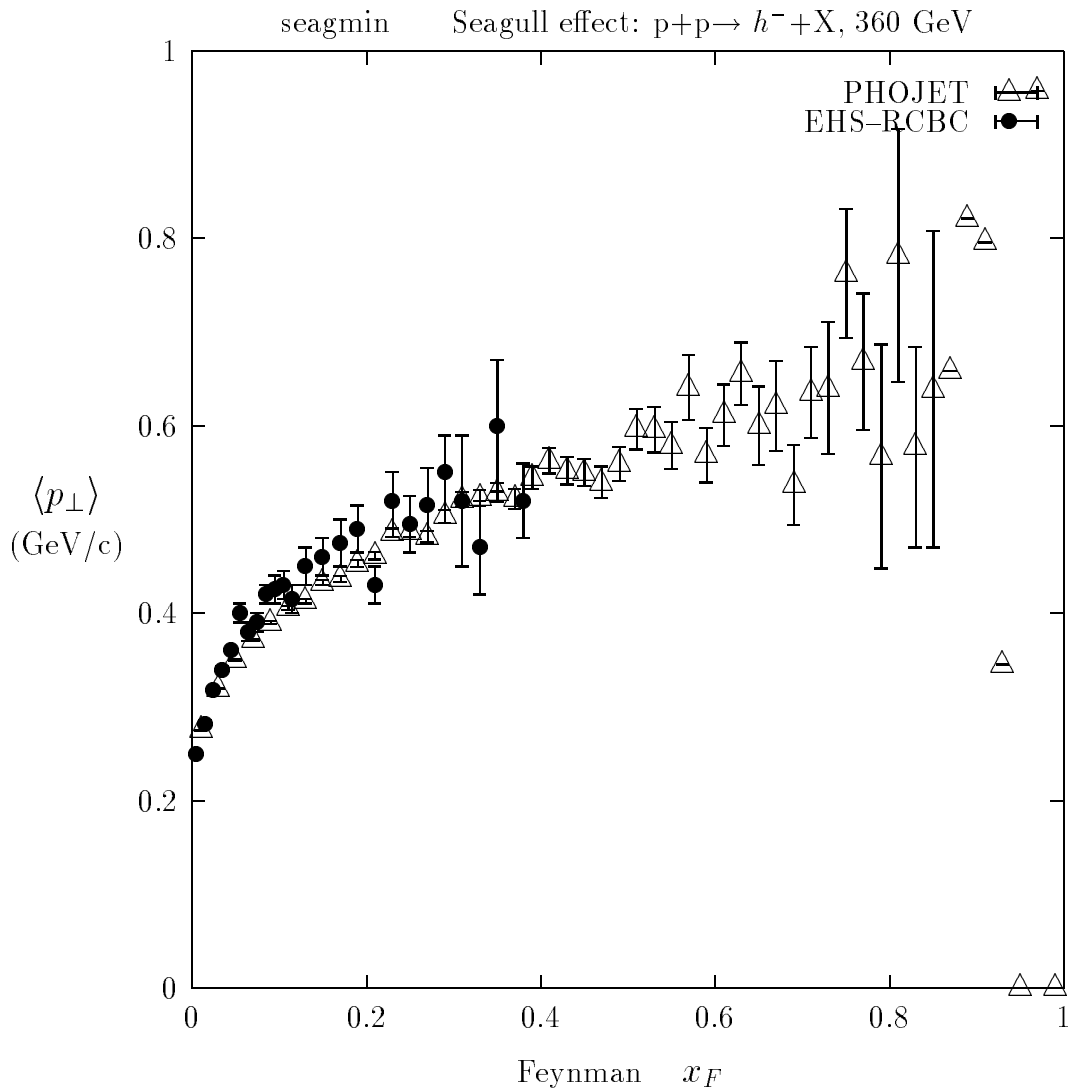

\caption{
Comparison of the seagull effect in the reaction p+p$\rightarrow
h^-$ +X at 360 GeV. The data are from the EHS-RCBC
Collaboration \protect\cite{Baillyseagull}. The calculation uses the
Dual Parton Model {\sc phojet}.
\label{seagmin}
}
\end{figure}

The seagull-effect, where one plots the average transverse
momentum $\langle p_{\perp}\rangle $ as function of Feynman $x_F$,
demonstrates clearly, that the distributions in transverse
momentum and longitudinal momentum are correlated in a
nontrivial way. There is no factorization between the transverse
momentum and longitudinal momentum distributions, as it is often
assumed in oversimplified models of hadron production. In Fig.
\ref{seagmin} we compare the seagull effect calculated with
{\sc phojet} to data
measured in $p_{lab}$ = 360 GeV/c proton-proton collisions
 \cite{Baillyseagull}.

\subsection{\bf Photon-hadron collisions}

The model for photon-hadron collisions is studied in full
detail  by Engel
\cite{engelgp}.  Here we present only some of this material in
order to make the present paper self-contained.

\begin{figure}
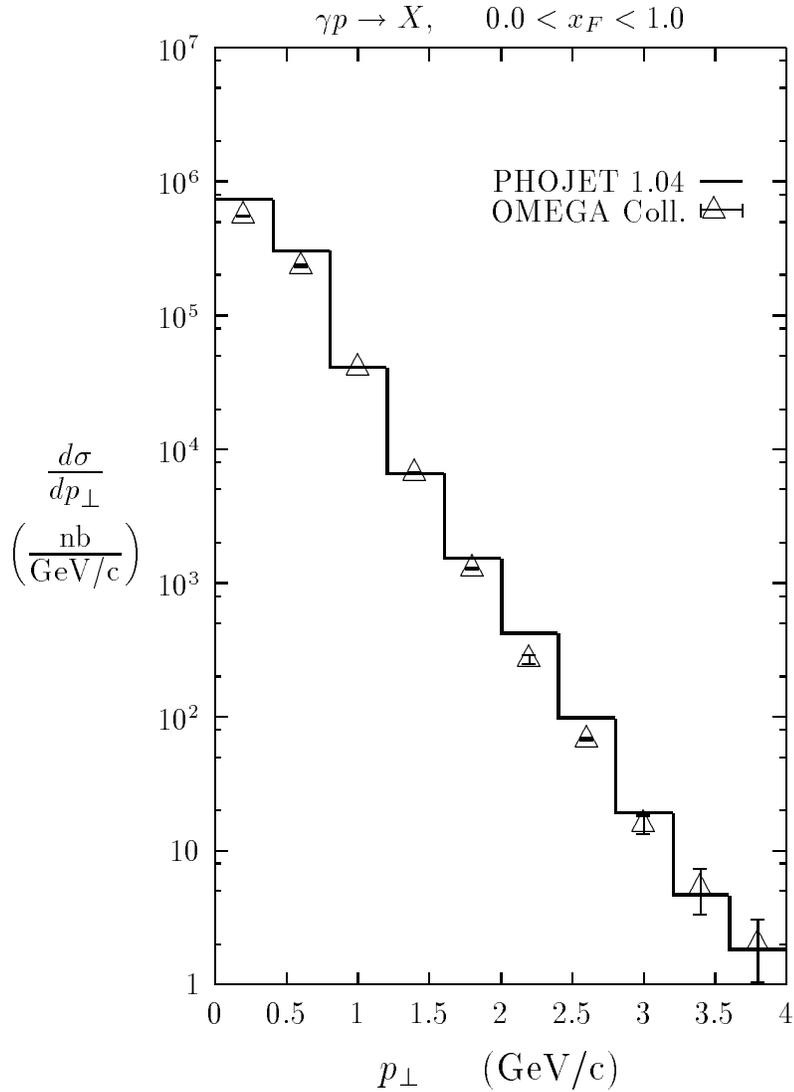

\caption{
Comparison of the transverse momentum distribution of charged
hadrons with $0 \le x_F \le 1$ as calculated with {\sc phojet}
at the average photon
energy of 140 GeV with data.  The data are from the OMEGA Collaboration
\protect\cite{Apsimon89a} measured with a tagged photon beam in the energy
band 110-170 GeV.
\label{omgp-pt}
}
\end{figure}

In Fig.~\ref{omgp-pt} we compare the transverse momentum distribution of
charged hadrons
 calculated with {\sc phojet} for photon-proton collisions
 with the photon lab. energy $E_{\gamma}$ = 140 GeV with data from
 the OMEGA Collaboration
\cite{Apsimon89a}. The data were obtained with a tagged photon
beam in the energy band 110-170 GeV. The agreement of both
distributions is excellent, at the largest $p_{\perp}$ we see
already the influence of direct and resolved hard collisions.

\begin{figure}
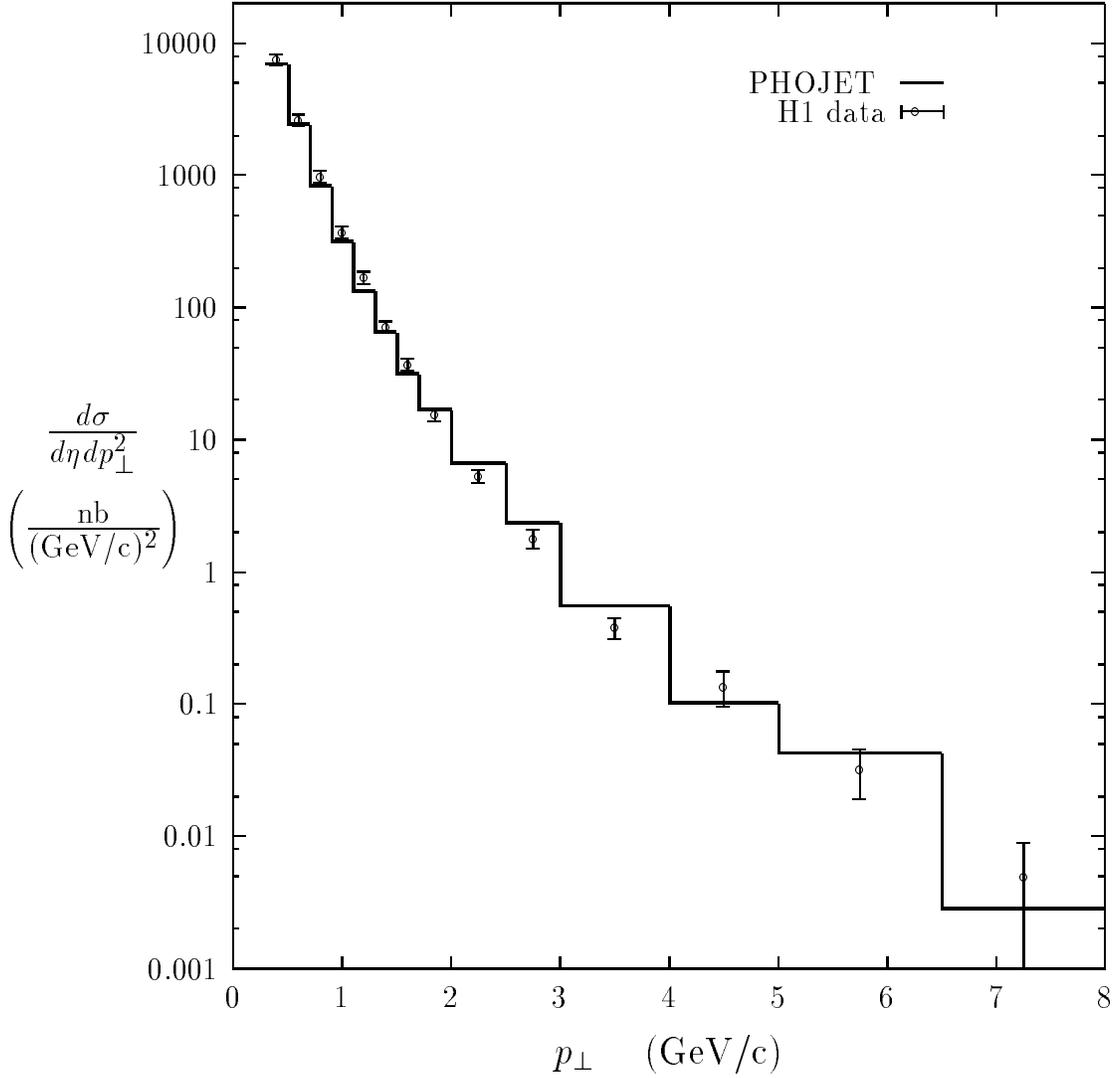

\caption{
Comparison of the transverse momentum distribution of charged
hadrons produced  by quasi-real photons in
electron-proton collisions at HERA \protect\cite{Abt94a} with the
{\sc phojet} calculation
\protect\cite{Engel94c}. The data and the calculation is in the
pseudorapidity region -1.5
\protect$\leq \eta_{\mbox{\scriptsize \ lab}} \leq$ 1.5.
\label{ggincpt}
}
\end{figure}

\begin{figure}
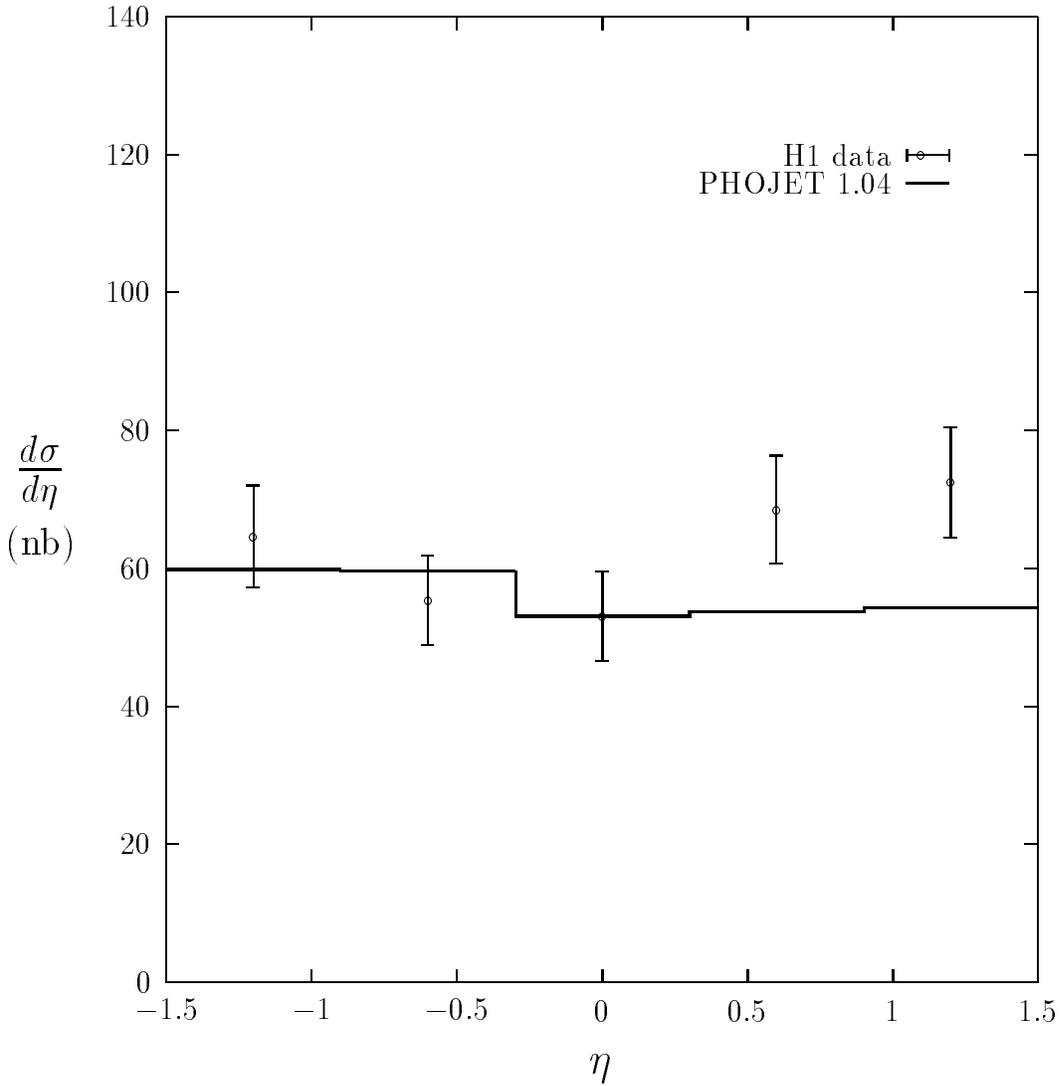

\caption{
Comparison of the inclusive pseudorapidity cross section of charged
hadrons produced  by quasi-real photons in
electron-proton collisions at HERA \protect\cite{Abt94a} with the
{\sc phojet} calculation
\protect\cite{Engel94c}.
\label{dsdeta}
}
\end{figure}

In Figs.
\ref{ggincpt} and
\ref{dsdeta} we compare {\sc phojet} results with data from the HERA
electron-proton collider
\cite{Abt94a}. In Fig.~\ref{ggincpt} we find a very good
agreement with the transverse momentum distribution of charged
hadrons. If we compare to proton-proton collisions
(Fig.~\ref{ua1dndpt}) we observe, that the influence of hard
collisions in photon-proton collisions is more prominent than
in proton-proton collisions. In Fig.~\ref{dsdeta} we find a
good agreement of the inclusive charged hadron production cross
section as function of the pseudorapidity. We stress, that the data
as well as the model show a flat pseudorapidity distribution.

\section{Comparison of minimum-bias hadron production in
hadron-hadron, photon-hadron and
photon-photon collisions}

In this Section we compare the model predictions for
inelastic hadron production  in
proton-proton, photon-proton and photon-photon collisions at fixed
center-of-mass (CMS) energies $\sqrt s$.
This is the usual way to present data on
hadron-hadron collisions at colliders or in fixed target
experiments. Collisions of quasi-real photons with protons at
electron-proton colliders and photon-photon collisions at
electron-positron colliders are not at fixed photon-proton or
photon-photon energy. We will discuss photon-photon collisions
at electron-positron colliders in the next Section. If we
consider  the three reaction channels at fixed energy, then only
in order to find the characteristic differences and similarities
between the three types of collisions. Since elastic
hadron-hadron collisions usually are excluded studying
inclusive secondary distributions,  again, in order to find the
similarities, we also exclude in the
reactions with photons  the corresponding quasi-elastic
diffractive channel i.e. $\gamma + \gamma \rightarrow V + V$,
($V = \rho , \omega ,\phi$)
but we include  all the other diffractive processes.

\begin{figure}
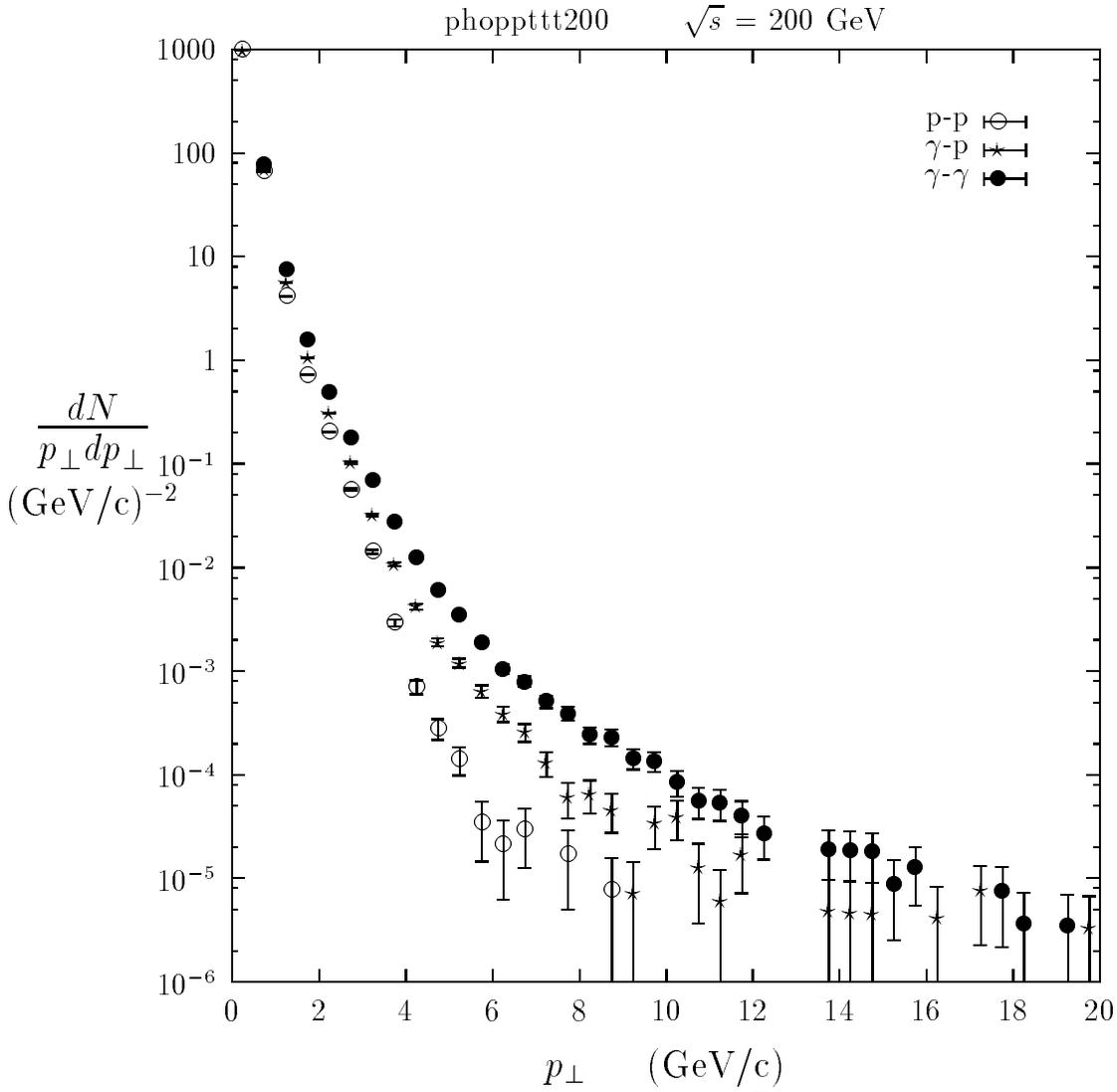

\caption{
We compare at the collision energy $\protect\sqrt s$ = 200 GeV the
transverse momentum distribution in invariant form for all
charged hadrons produced in proton-proton, photon-proton and
photon-photon collisions. The calculation was done with
{\sc phojet} for inelastic collisions, excluding in
photon-proton collisions the $V$-$p$ and in photon-photon
collisions the $V$-$V$ diffractive production of vector mesons
$V = \protect\rho$, $\protect\omega$ and $\protect\phi$.
\label{phoppttt200}
}
\end{figure}

We start with the distribution, where the three channels differ
most strongly, the transverse momentum distribution
$dN/p_{\perp}dp_{\perp}$, see Fig.~\ref{phoppttt200}. The
difference is striking. The fraction of hard interactions in minimum bias
interactions rises from
proton-proton collisions over photon-proton collisions to
photon-photon collisions. The reason for this is the direct
photon interaction and the fact, that the photon structure
function is considerably harder than the proton structure
function. In photon-photon collisions it is easy to observe
already with moderate statistics hadrons with transverse
momenta close to the kinematic limit.

However, these differences in the hard scattering do  not
strongly
influence  such average properties of the collision as average
multiplicities or even average transverse momenta. This can be
seen from Table 1, where we collect some average quantities
characterizing  nondiffractive
proton-proton, proton-photon and photon-photon
collisions at CMS energies between 10 and 200 GeV. The total
and charged multiplicities at all energies are rather near to
each other in all channels. Probably the differences in the
multiplicities of newly produced hadrons like $\pi^-$ and $\bar
p$ are more significant, we find them at all energies rising from
$p$-$p$ over $\gamma$-$p$ to $\gamma$-$\gamma$ collisions.
Also the average transverse momenta rise as expected from
$p$-$p$ over $\gamma$-$p$ to $\gamma$-$\gamma$.
In Table 1 we give also the
number of soft pomerons $\langle n\rangle_{\mbox{\scriptsize softch.}}$
and the number of hard
pomerons $\langle n\rangle_{\mbox{\scriptsize minijets}}$
contributing in average to the
hadronic final state.
The numbers given are obtained after kinematical
corrections due to energy momentum conservation,
not the numbers obtained from the unitarization
step
(it turns out, especially at low energies, that for
kinematical reasons not all sampled   pomeron cuts
can be generated  in the
Monte Carlo).
At low energy, where the number of minijets is very small,
we find the number of cut soft pomerons to be the same in all three
channels.
The number of cut hard pomerons rises at all energies from
$p$-$p$ over $\gamma$-$p$ to $\gamma$-$\gamma$. We see also, that
the numbers of cut soft and cut hard pomerons are correlated, at high
energies the number of cut soft pomerons decreases from
$p$-$p$ over $\gamma$-$p$ to $\gamma$-$\gamma$.

In Fig.~\ref{phoxr200} we compare the longitudinal momentum
distributions in the form of $x_rdN/dx_r$ for the three
channels. $x_r = (p_\|/|p_\||)2E/\sqrt s$ is the so called
radial scaling
variable, very similar to the Feynman variable $x_F = 2p_\|/\sqrt
s$. Significant differences between the three channels are only
found in the region near $x_r$ = 1 or -1. The reason is the
single diffractive component, which in the $p$-$p$ and $\gamma$-$p$
case leads to the diffractive protons, which are obviously
missing in $\gamma$-$\gamma$ collisions.

\begin{figure}
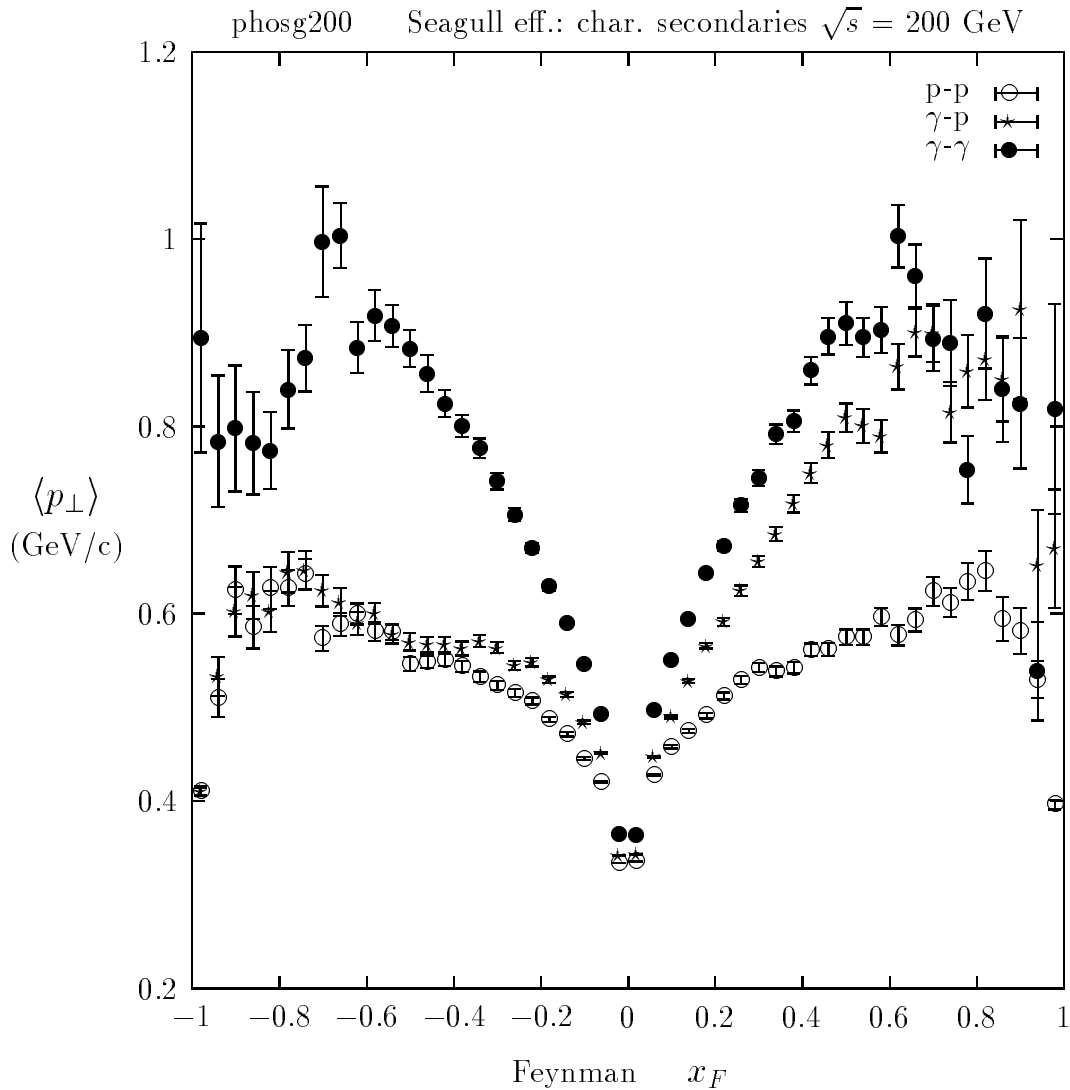

\caption{
We compare at the collision energy $\protect\sqrt s$ = 200 GeV the
 average transverse momentum of
charged hadrons
produced in proton-proton, photon-proton and
photon-photon collisions
as function of the Feynman $x_F$ variable
(seagull effect). The calculation was done with
{\sc phojet} for inelastic collisions, excluding in
photon-proton collisions the $V$-$p$ and in photon-photon
collisions the $V$-$V$ diffractive contributions.
\label{phosg200}
}
\end{figure}

In Fig.~\ref{phosg200} we compare the seagull effect in the
three channels. We find due to the direct processes in
photon-photon collisions, the rise of $\langle p_{\perp}\rangle $
with rising
Feynman $x_F$ is more prominent in photon-photon collisions
than in proton-proton collisions.
In photon-proton interactions we find backwards, in the proton
fragmentation region, agreement with the $p$-$p$ collision and in
forward direction, the photon fragmentation region, agreement
with photon-photon collisions.

In Figs. \ref{phoetet} and \ref{phoetel} we plot first
logarithmically and then linearly the transverse energy
distribution $dE_{\perp}/d\eta$. Roughly, these distributions should be
equivalent to the pseudorapidity distribution $dN/d\eta$
multiplied with the average transverse energy per particle. We
observe characteristic differences, which can be understood from
the features already discussed. The transverse energy
distribution is wider in photon-photon collisions than in
proton-proton collisions. The transverse energy distribution
rises at all $\eta$ from $p$-$p$ over $\gamma$-$p$ to
$\gamma$-$\gamma$. For $\gamma$-$p$ the distribution agrees
backwards with $p$-$p$ and forwards with $\gamma$-$\gamma$.

For $p$-$p$ or $\bar p$-$p$
collisions the transverse energy distribution at
$\eta$ = 0 is known from ISR  experiments and experiments at
the CERN collider. The values calculated with {\sc phojet}
agree  well with these measurements. At HERA it was found,
that practically the same transverse energy at $\eta$ = 0 is
found like in $p$-$p$ interactions and in collisions of real or
virtual photons with protons \cite{Aid95a}.
This observation agrees well with  our results
in Figs. \ref{phoetet} and \ref{phoetel}.

\begin{figure}
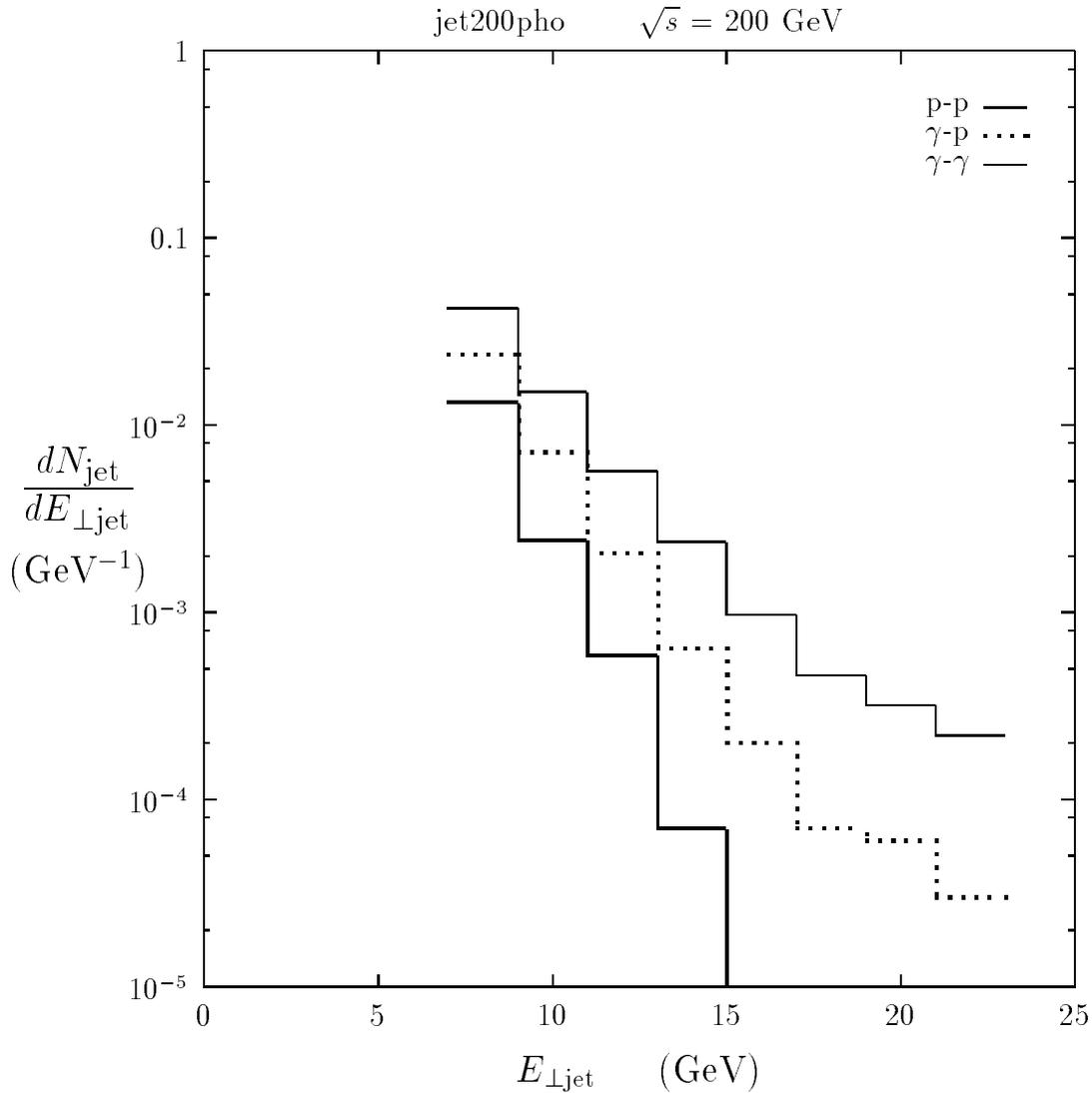

\caption{
We compare at the collision energy $\protect\sqrt s$ = 200 GeV the
transverse energy distribution for hadronic jets (identified
using a jet-finding algorithm)
 produced in proton-proton, photon-proton and
photon-photon collisions. The calculation was done with
{\sc phojet}.
\label{jet200pho}
}
\end{figure}

Finally in Fig.~\ref{jet200pho} we plot the transverse energy
distributions
$dN_{\mbox{\scriptsize jet}}/dE_{\perp \mbox{\scriptsize jet}}$
of jets found  in the Monte
Carlo events from {\sc phojet}.  The jets are searched on hadron
level using a cone jet finding algorithm with the cone radius
R=1.  We find again: high $E_{\perp \mbox{\scriptsize jet}}$
jets are more prominent in
$\gamma$-$\gamma$ collisions.

\section{ Hadron production in photon-photon collisions at
present and future electron-positron colliders}

\subsection{\bf Photon flux calculation}

\subsubsection{Bremsstrahlung}

The flux of quasi-real photons is calculated using
the equivalent photon approximation (improved Weizs\"acker-Williams spectrum
\cite{Weizsacker34,Williams34,Frixione93}).
Within this approximation
the $ep \rightarrow eX$ photoproduction cross section is given by
\begin{equation}
\frac{d\sigma_{ep}}{dy} = f_{\gamma,e}(y)\ \sigma_{\gamma p}(s,0)
\end{equation}
with
\begin{equation}
f_{\gamma,e}(y) = \frac{\alpha_{\mbox{\scriptsize em}}}{2 \pi}
\left[ \frac{1+(1-y)^2}{y}
\mbox{ln} \frac{P^2_{\mbox{\scriptsize max}}}{P^2_{\mbox{\scriptsize min}}}
- 2 m_e^2 y \left(
\frac{1}{P_{\mbox{\scriptsize min}}^2}
-\frac{1}{P_{\mbox{\scriptsize max}}^2}\right)
\right].
\end{equation}
Here, $y$ and $P^2=-p_\gamma^2$ denote
the energy fraction taken by the photon from the electron
and the photon virtuality.
$\alpha_{\mbox{\scriptsize em}}$ is the fine structure constant.
Taking the kinematic limit
$P_{\mbox{\scriptsize min,kin}}^2$ as lowest photon virtuality allowed
one gets with the electron mass $m_e$ and
\begin{equation}
 P^2_{\mbox{\scriptsize min,kin}} = \frac{m^2_e y^2}{1-y}
\end{equation}
the spectrum of quasi-real photons
\begin{equation}
f_{\gamma,e}(y) = \frac{\alpha_{\mbox{\scriptsize em}}}{2 \pi}
\left( \frac{1+(1-y)^2}{y} \mbox{ln}
\frac{(1-y)P^2_{\mbox{\scriptsize max}}}{m_e^2 y^2}
- \frac{2(1-y)}{y} \right).
\end{equation}

A similar expression holds for $e e \rightarrow e e + X$ scattering
involving quasi-real photons only
\begin{equation}
\frac{d^2\sigma_{ep}}{dy_1dy_2} = f_{\gamma,e}(y_1)\; f_{\gamma,e}(y_2)
\ \sigma_{\gamma \gamma}(s,0).
\end{equation}

\subsubsection{Beamstrahlung}

In case of Gaussian beams, the effective beamstrahlung spectrum
has been estimated by Chen et.al.\ \cite{Chen94a}. The dependence of this
spectrum on the particle-bunch parameters can be expressed by the
beamstrahlung parameter $Y$:
\begin{equation}
Y=\frac{5 r_e^2 E N_e}{6 \alpha_{\mbox{\scriptsize em}}
\sigma_z (\sigma_x + \sigma_y) m_e}.
\end{equation}
Here, $E$ denotes the beam energy,
$N_e$ is the number of electrons or positrons in a bunch, $\sigma_x$
and $\sigma_y$ are the transverse bunch dimensions, and
$r_e=2.818 \cdot 10^{-12}$mm is the classical electron radius.
The beamstrahlung spectrum is approximated by \cite{Chen94a,Drees93c}
\begin{eqnarray}
f_{\gamma,e}^{\mbox{\scriptsize beam}}\left(y\right) &=&
\frac{\kappa^{1/3}}{\Gamma(1/3)} y^{-2/3} \left(1-y)\right)^{-1/3}
e^{-\kappa y/(1-y)} \nonumber\\
&\cdot& \bigg\{\frac{1-w}{\tilde{g}(y)}
\left[1-\frac{1}{\tilde{g}(y)N_\gamma}\left(1-e^{-N_\gamma\tilde{g}(y)}
\right)\right]\nonumber\\
&+& w\left[1-\frac{1}{N_\gamma}\left(1-e^{-N_\gamma}\right)\right]\bigg\},
\end{eqnarray}
with
\begin{equation}
\tilde{g}(y)=1-\frac{1}{2} \left(1-y\right)^{2/3}
\left[1-y+(1+y)\sqrt{1+Y^{2/3}}\right]
\end{equation}
and $\kappa=2/(3Y)$, $w=1/(6\sqrt{\kappa})$.
The average number
of photons $N_\gamma$ emitted per electron is given by
\begin{equation}
N_\gamma  = \frac{5\alpha_{\mbox{\scriptsize em}}^{2}\sigma_z m_e}{2r_e E}
            \frac{Y}{\sqrt{1+Y^{2/3}}}.
\end{equation}
\subsubsection{Photonemission by laser-backscattering}

Depending on the polarization of the laser light, various photon spectra
can be produced \cite{Ginzburg83a,Ginzburg84a,Telnov90a}.
Here we consider only the case of
unpolarized laser radiation. Furthermore, we assume that the laser
frequency is chosen to be below the pair-creation threshold at the
optimal point.
Then, the spectrum of backscattered photons
can be written as \cite{Drees93c}
\begin{equation}
f_{\gamma,e}^{\mbox{\scriptsize laser}}(y)=
\frac{-0.544\;y^3 + 2.17\;y^2 -2.63\;y + 1.09}{(1-y)^2}\
\Theta(0.828 - y).
\end{equation}
\subsection{\bf Hadron production in photon-photon collisions at
 electron-positron linear
colliders}

There are at present several projects
for electron-positron
linear colliders under active study. Here we pick out for our calculations
only one of these projects, the TESLA linear collider. Details about the
most recent TESLA project were given by Brinkmann \cite{brinkmann}.

\begin{figure}
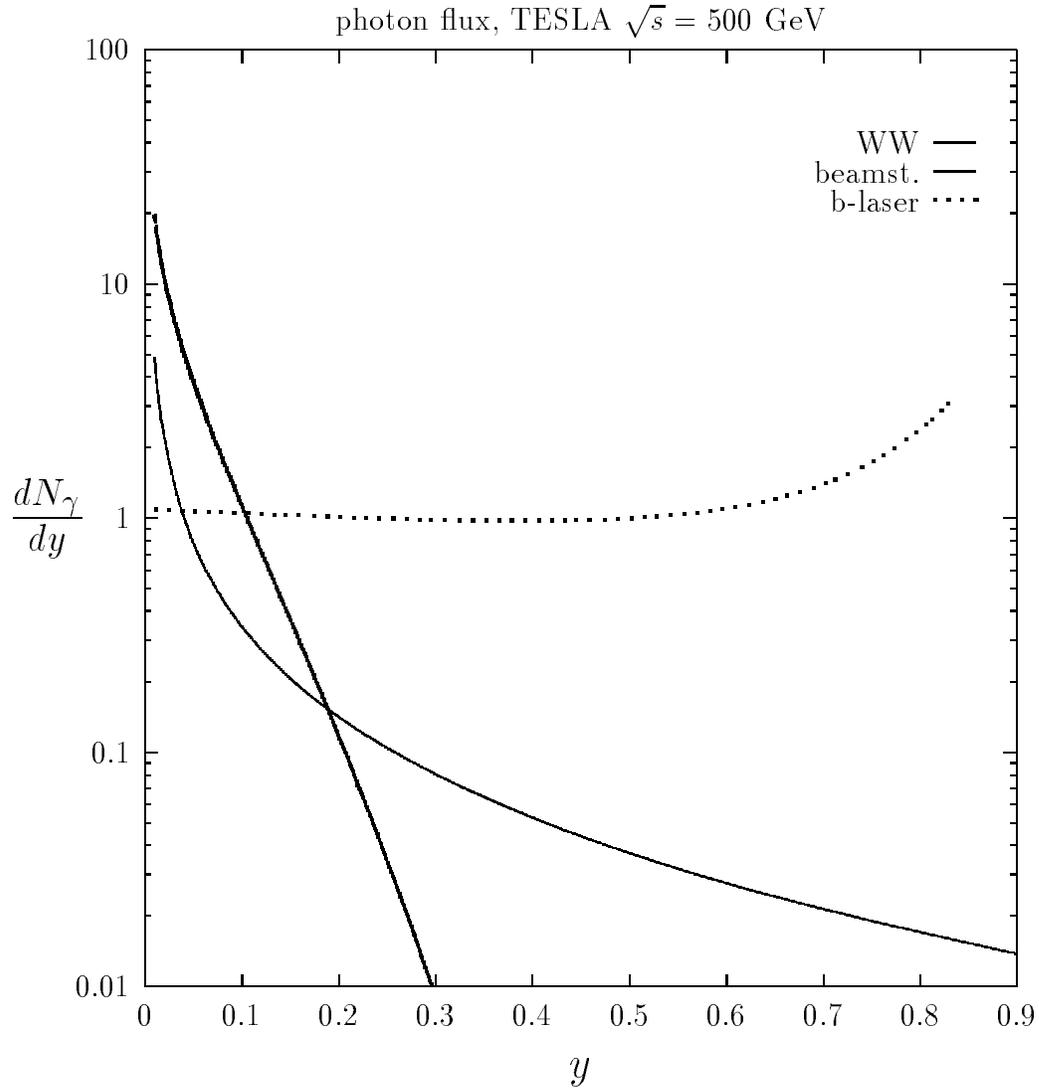

\caption{
Photon fluxes at a $\protect\sqrt s$ = 500 GeV linear collider TESLA
\protect\cite{brinkmann}.
Given are the improved Weizs\"acker-Williams spectrum,
the beamstrahlung spectrum using the bunch parameters $N_e$ =
1.8 10$^{10}$, $\sigma_x$ = 598 nm, $\sigma_y$ = 6.5 nm  and
$\sigma_z$ = 0.5 mm \protect\cite{brinkmann}, and a backscattered laser
spectrum.
\label{flux1}
}
\end{figure}

Using the formulae discussed in the last subsection we plot in
Fig.~\ref{flux1} the photon spectra according to the
equivalent photon approximation, the beamstrahlung spectrum using
the bunch parameters \cite{brinkmann} as given in the caption of
Fig.~\ref{flux1}
 and a backscattered laser spectrum. The photon virtuality
was restricted to $P^2 \le 0.01$ GeV$^2$/c$^2$. In Table 2 we give
the average photon-photon energies and the weight factors in
$\mu$b for the three photon spectra and two energies. From
Fig.~\ref{flux1} and Table 2 we see, that the beamstrahlung
spectrum of the TESLA project is the softest of the three photon
spectra, the backscattered laser spectrum is the hardest.

Of
course, in the case of a linear collider we will always have to
consider for background problems the
superposition of the beamstrahlung spectrum and the
equivalent photon spectrum.

A reasonable lowest energy for collisions to be sampled using
{\sc phojet} is $\sqrt s_{\gamma \gamma}$ = 5 GeV ({\sc phojet} would
however run without problems even down to $\sqrt s_{\gamma
\gamma}$ = 2 GeV).
Therefore, for all applications in this and the following
subsection we always cut the photon spectra at small $y$ in such
a way, that this lower energy cut-off  is respected.

\begin{figure}
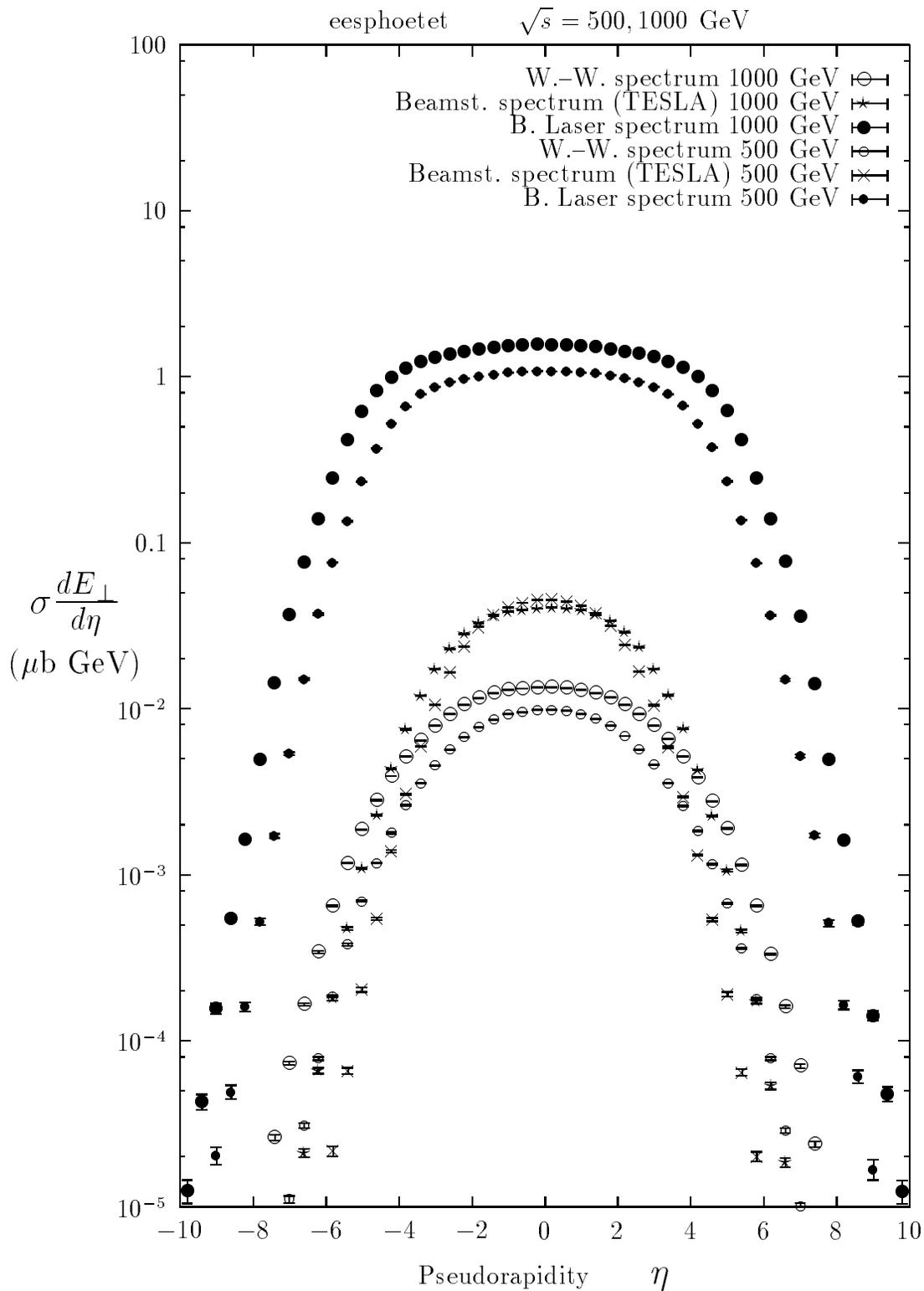

\caption{
Cross section weighted transverse energy distributions
$\sigma dE_{\perp}/d\eta$ measured in $\mu$b~GeV at
the $\protect\sqrt s$ = 500 and
1000 GeV TESLA linear colliders \protect\cite{brinkmann}.
Given are the
distributions for the Weizs\"acker-Williams photon spectrum,
the beamstrahlung spectrum and a backscattered laser spectrum.
\label{eesphoetet}
}
\end{figure}

\begin{figure}
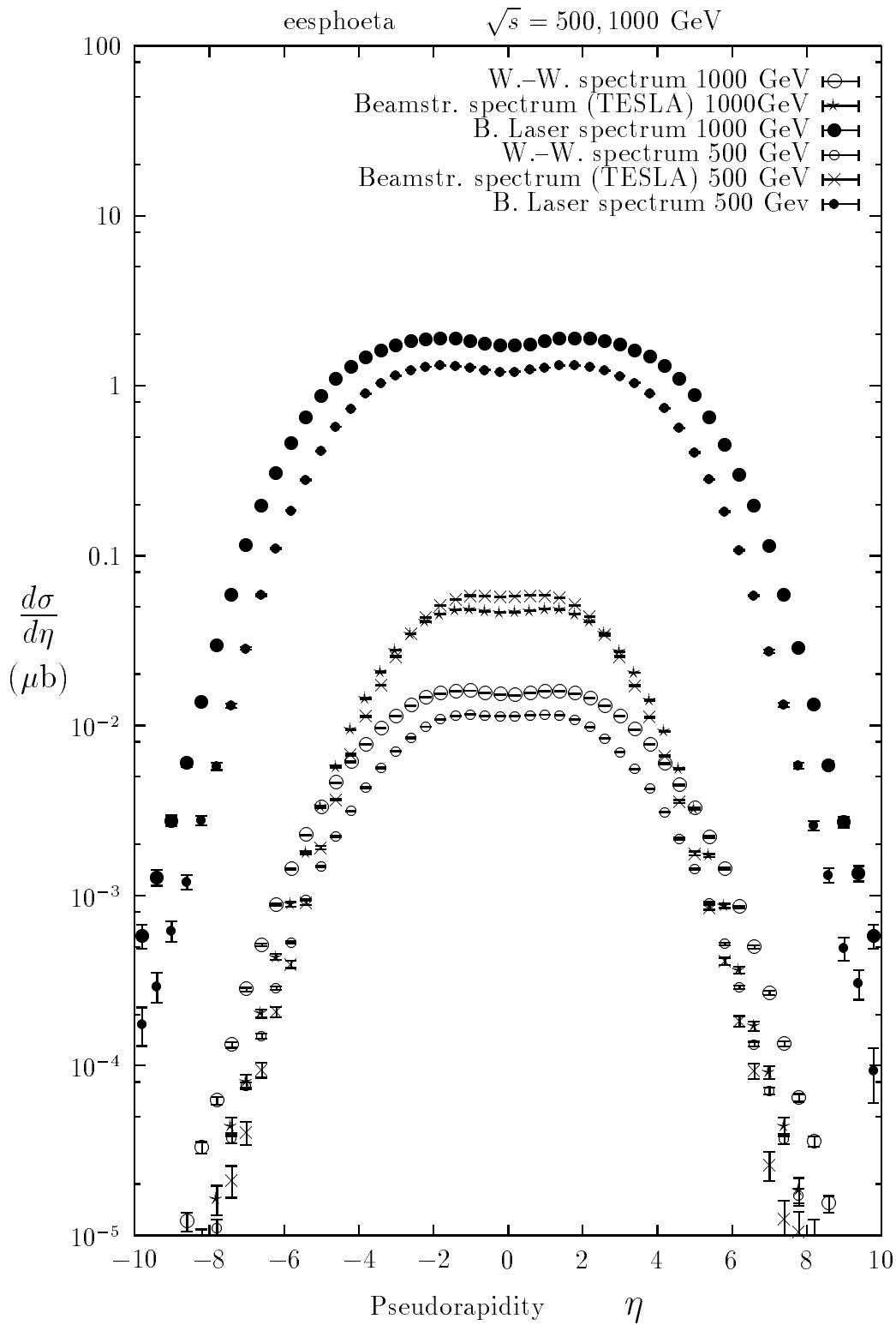

\caption{
Pseudorapidity cross sections $d\sigma/d\eta$, measured in $\mu$b
per pseudorapidity unit, at
the $\protect\sqrt s$ = 500 and
1000 GeV TESLA linear colliders \protect\cite{brinkmann}.
Given are the
distributions for the Weizs\"acker-Williams photon spectrum,
the beamstrahlung spectrum and a backscattered laser spectrum.
\label{eesphoeta}
}
\end{figure}

In Figs. \ref{eesphoetet} and \ref{eesphoeta} we plot the
cross sections $\sigma dE_{\perp}/d\eta$ for the transverse energy as
function of pseudorapidity and $d\sigma/d\eta$ for the
charged hadron production as function of pseudorapidity.
It is clearly visible, that the backscattered laser spectrum is rather
hard and has the highest weight. The beamstrahlung spectrum and
the equivalent photon spectrum are rather comparable, the
former has the higher weight, the latter is the harder of these
two.

\begin{figure}
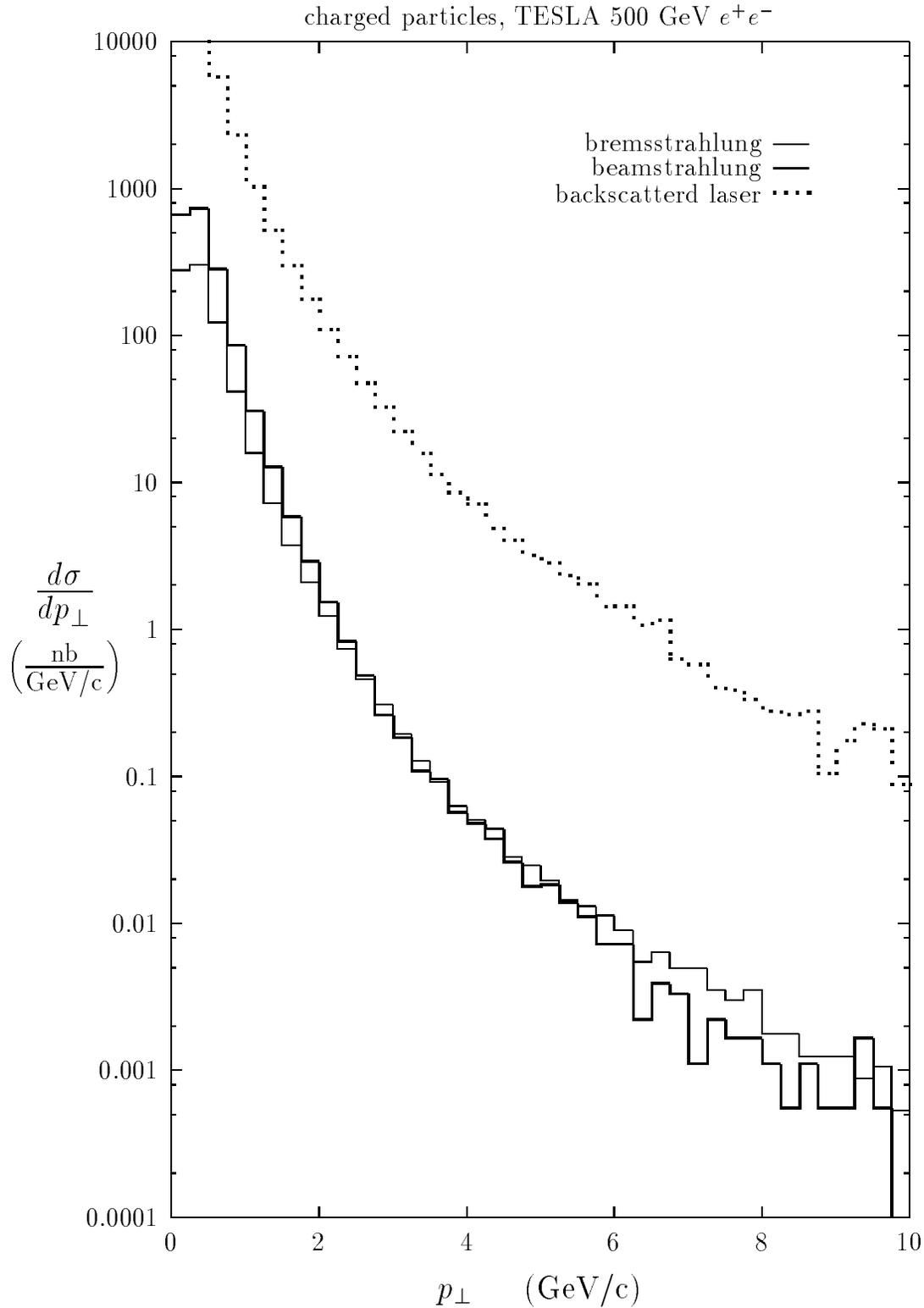

\caption{
Transverse momentum cross sections $d\sigma/dp_{\perp}$
at the $\protect\sqrt s$ = 500 GeV TESLA linear collider
\protect\cite{brinkmann}.
Given are the
distributions for the Weizs\"acker-Williams photon spectrum,
the beamstrahlung spectrum and a backscattered laser spectrum.
Please note that the $p_{\perp}$ distributions for the
Weizs\"acker-Williams spectrum and the beamstrahlung spectrum
cross. At low $p_{\perp}$ the beamstrahlung dominates, at high
$p_{\perp}$  the W.-W. spectrum dominates.
\label{eesphopptt500}
}
\end{figure}

The same differences between the three photon spectra are
visible in the cross sections $d\sigma/dp_{\perp}$
for charged hadron production as
function of the transverse momentum
in Fig.~\ref{eesphopptt500} for the 500 GeV TESLA collider.

\subsection{\bf Hadron production in photon-photon collisions at
the LEP-II electron-positron collider}

We use throughout this section an electron-positron energy of
$\sqrt s$ = 175 GeV for the phase II of the LEP collider. Of course
here we have only to consider the equivalent  photon
spectrum. Assuming anti-tagging of the scattered electons and positrons we
restrict the photon virtuality to satisfy $P^2 \le 1.75$ GeV$^2$/c$^2$.

The average photon-photon energy at LEP-II with the given
electron-positron energy will be ${\sqrt s}_{\gamma\gamma}$ = 25 GeV.
In Table 3 we compare some average properties of the LEP-II
photon-photon collisions with photon-photon collisions at
${\sqrt s}_{\gamma\gamma}$ = 25 GeV. We find, that the average
properties are rather well represented by photon-photon
collisions at $\sqrt s$ = 25 GeV.

In the following we consider a calorimeter detector with the pseudorapidity
coverage $|\eta| \leq$ 2.1. Applying  a jet finding algorithm to
this detector we can
identify jets in the pseudorapidity range
$|\eta_{\mbox{\scriptsize jet}}| \leq$ 1.1.
For the jets we use generally a lower jet transverse energy
cutoff $E_{\perp \mbox{\scriptsize jet}} \geq$ 5GeV.

\begin{figure}
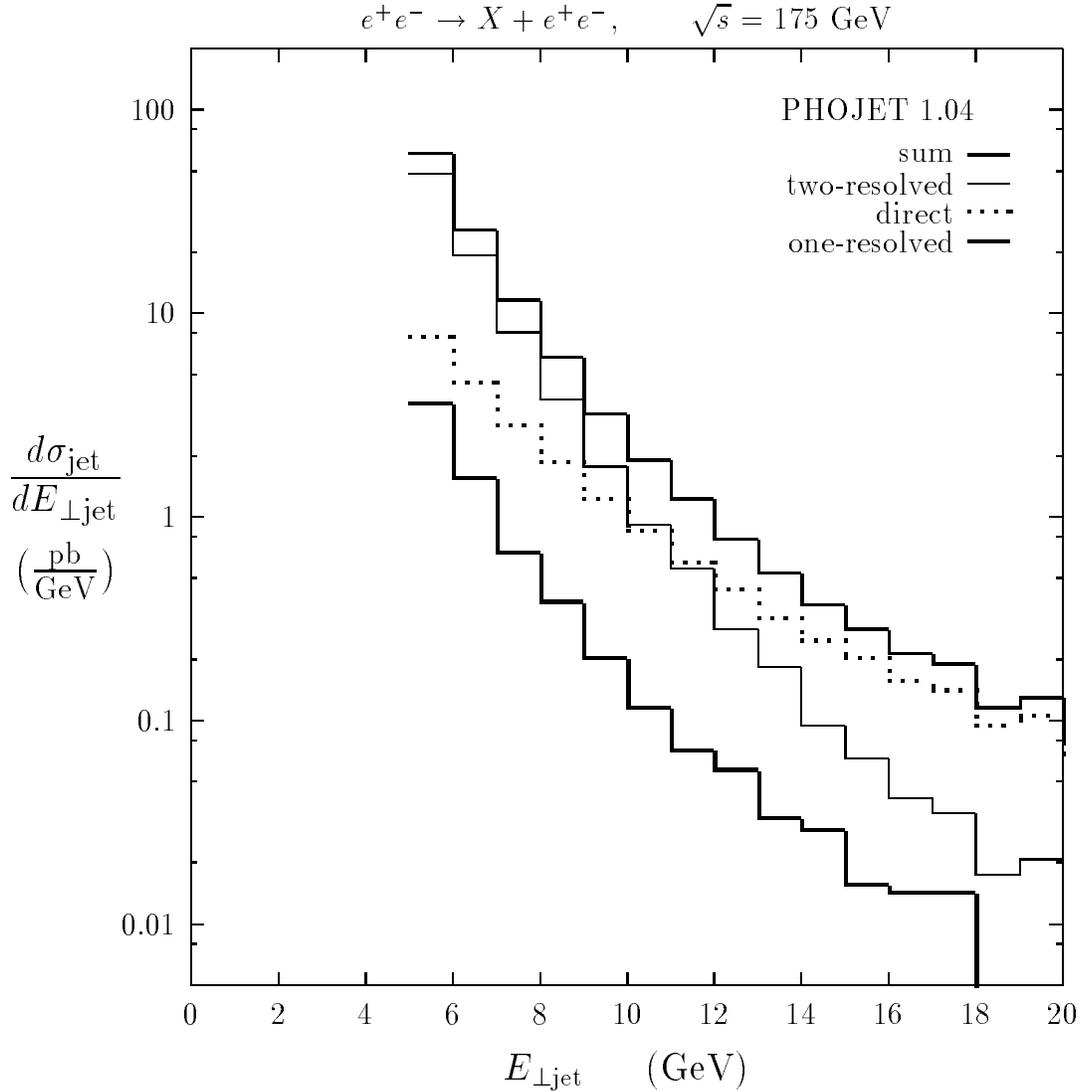

\caption{
Decomposition of the jet cross section
$d\sigma_{\mbox{\scriptsize jet}}/dE_{\perp \mbox{\scriptsize jet}}$
as function of the jet transverse energy.
The jets with
\protect$E_{\perp \mbox{\scriptsize jet}} \geq$ 5 GeV were found from
the {\sc phojet} events using a cone algorithm.
The jet acceptance region was restricted to
\protect$|\eta_{\mbox{\scriptsize jet}}| \leq$ 1.1.
\label{ggjetets}
}
\end{figure}

In Fig.~\ref{ggjetets} we give the jet transverse energy
distribution  in the  acceptance region given above
together with its
decomposition into the different hard interaction mechanisms. The
single resolved (equal to the single direct) contribution within
the acceptance region is surprisingly small. At small
$E_{\perp \mbox{\scriptsize jet}}$
the two resolved contribution dominates, at large
$E_{\perp \mbox{\scriptsize jet}}$ the direct contribution dominates.

\begin{figure}
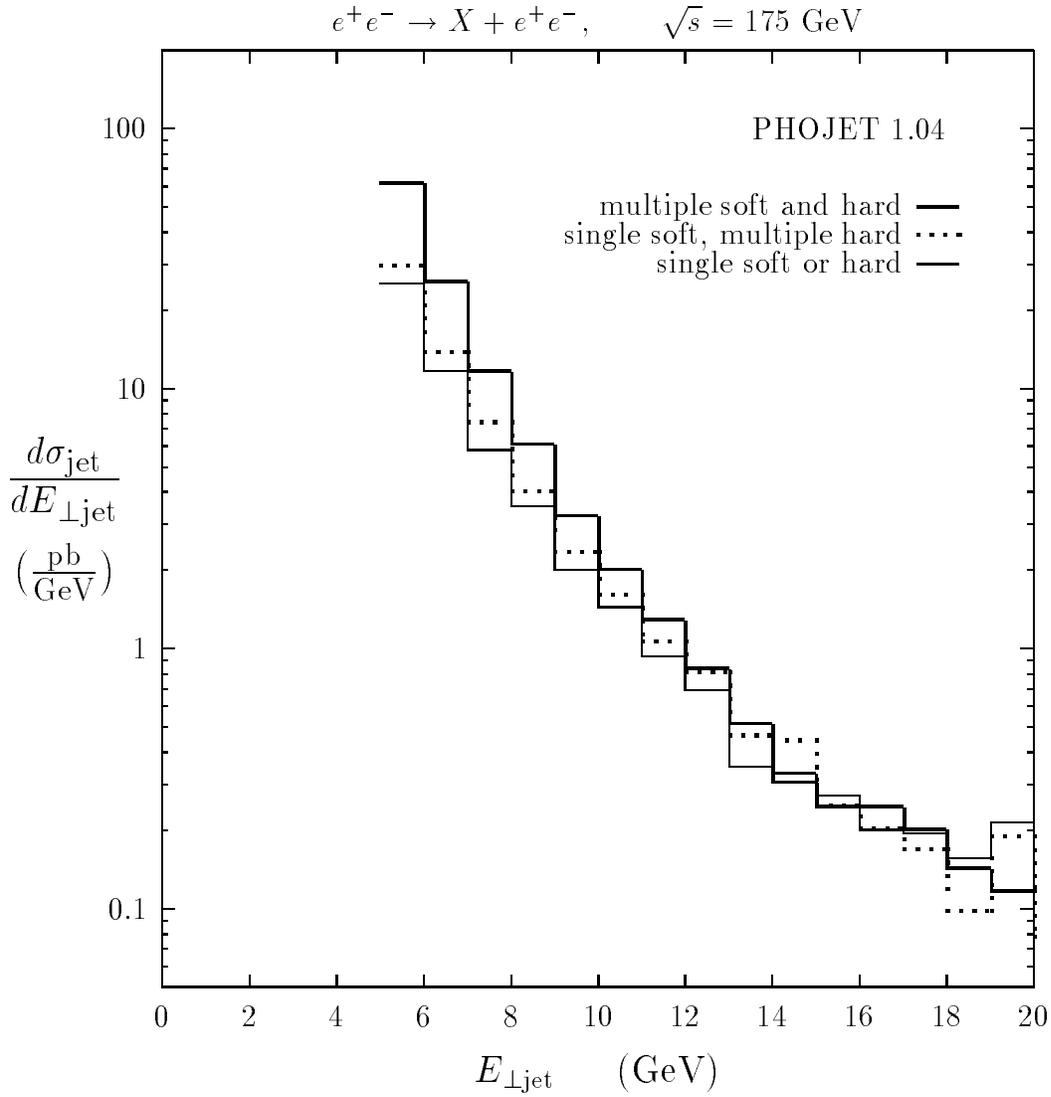

\caption{
Jet cross section
\protect$d\sigma_{\mbox{\scriptsize jet}}/dE_{\perp \mbox{\scriptsize jet}}$
as function of the jet transverse energy.
In order to study the influence of multiple soft and hard
interactions in the model, we consider restricted cases with
only single soft  or single hard collisions. The curve with
multiple soft, single hard collision coincides nearly with the
full curve.
The jets with
\protect$E_{\perp \mbox{\scriptsize jet}} \geq$ 5 GeV were found from
the {\sc phojet} events using a cone algorithm.
The jet acceptance region was restricted to
\protect$|\eta_\protect{\mbox{\scriptsize jet}}| \leq$ 1.1.
\label{ggjetetm}
}
\end{figure}

To study the influence of multiple soft and multiple hard
interactions in the model we plot in Fig.~\ref{ggjetetm} the jet
$E_{\perp}$ cross section together with further curves, where
multiple soft or multiple hard contributions or both are  rejected.
The arbitrary threshold between soft and hard partons in the
model is $p_{\perp}^{\mbox{\scriptsize cutoff}} =$ 3 GeV/c.
The multiple soft, one hard curve coincides
nearly with the total $E_{\perp \mbox{\scriptsize jet}}$ distribution.
With the average numbers
of minijets given in Table 3 the chance to have more than one
minijet in one event is  small. This is different for the
multiple soft interactions, with an average (see Table 3)
around 1.36 soft interactions per event, the chance to have multiple
soft chains  is rather high.

\begin{figure}
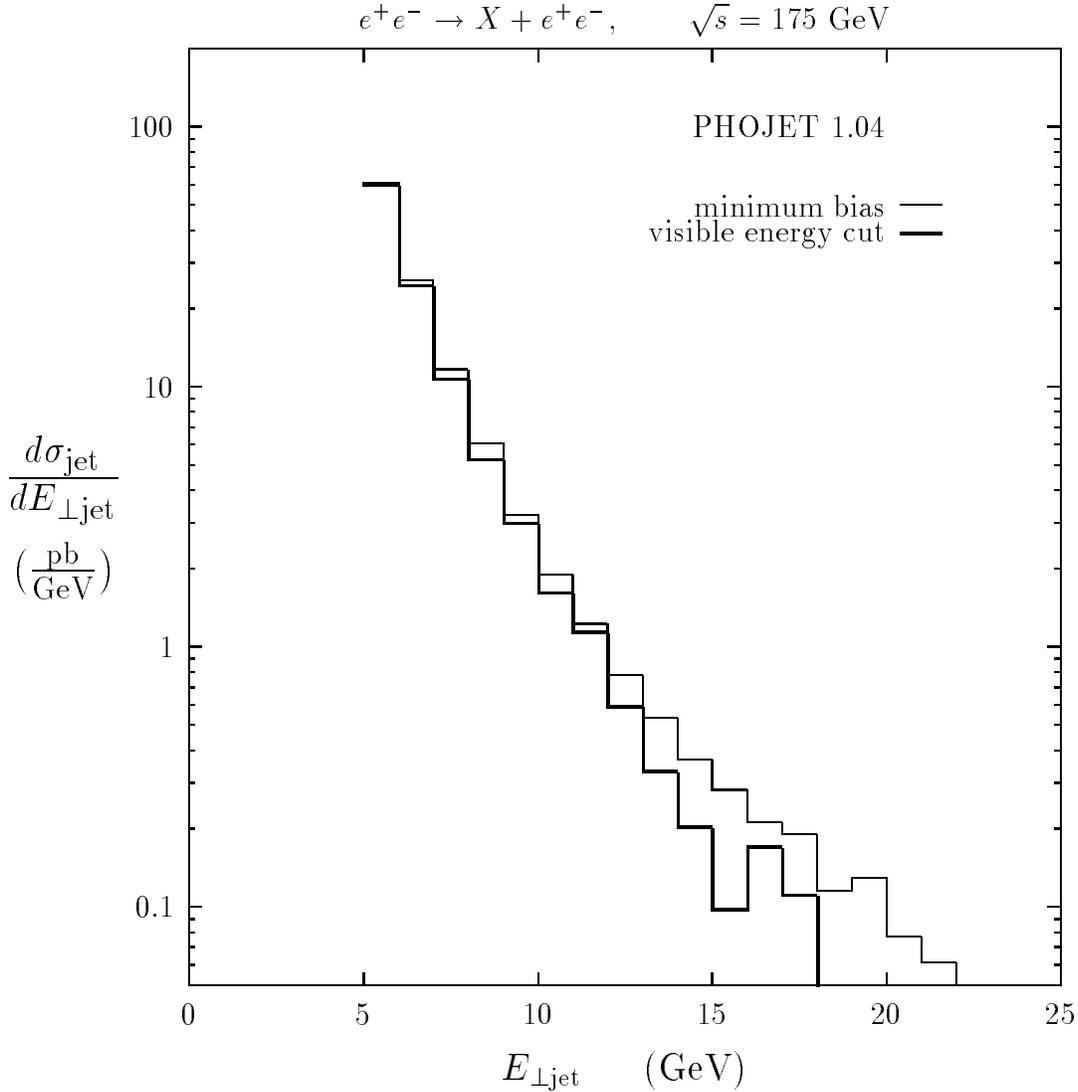

\caption{
Jet cross section
\protect$d\sigma_{\mbox{\scriptsize jet}}/dE_{\perp \mbox{\scriptsize jet}}$
as function of the transverse energy.
Here we give in addition the cross section with a visible energy
cut (inside a calorimeter with \protect$|\eta| \leq$ 2.1 the cut is
$E_{\mbox{\scriptsize visible}} \leq$ 50 GeV).
The jets with $E_{\perp \mbox{\scriptsize jet}} \geq$ 5 GeV were found from
the {\sc phojet} events using a cone algorithm.
The jet acceptance region was restricted to
$|\eta_{\mbox{\scriptsize jet}}| \leq$ 1.1.
\label{ggjetetv}
}
\end{figure}

\begin{figure}
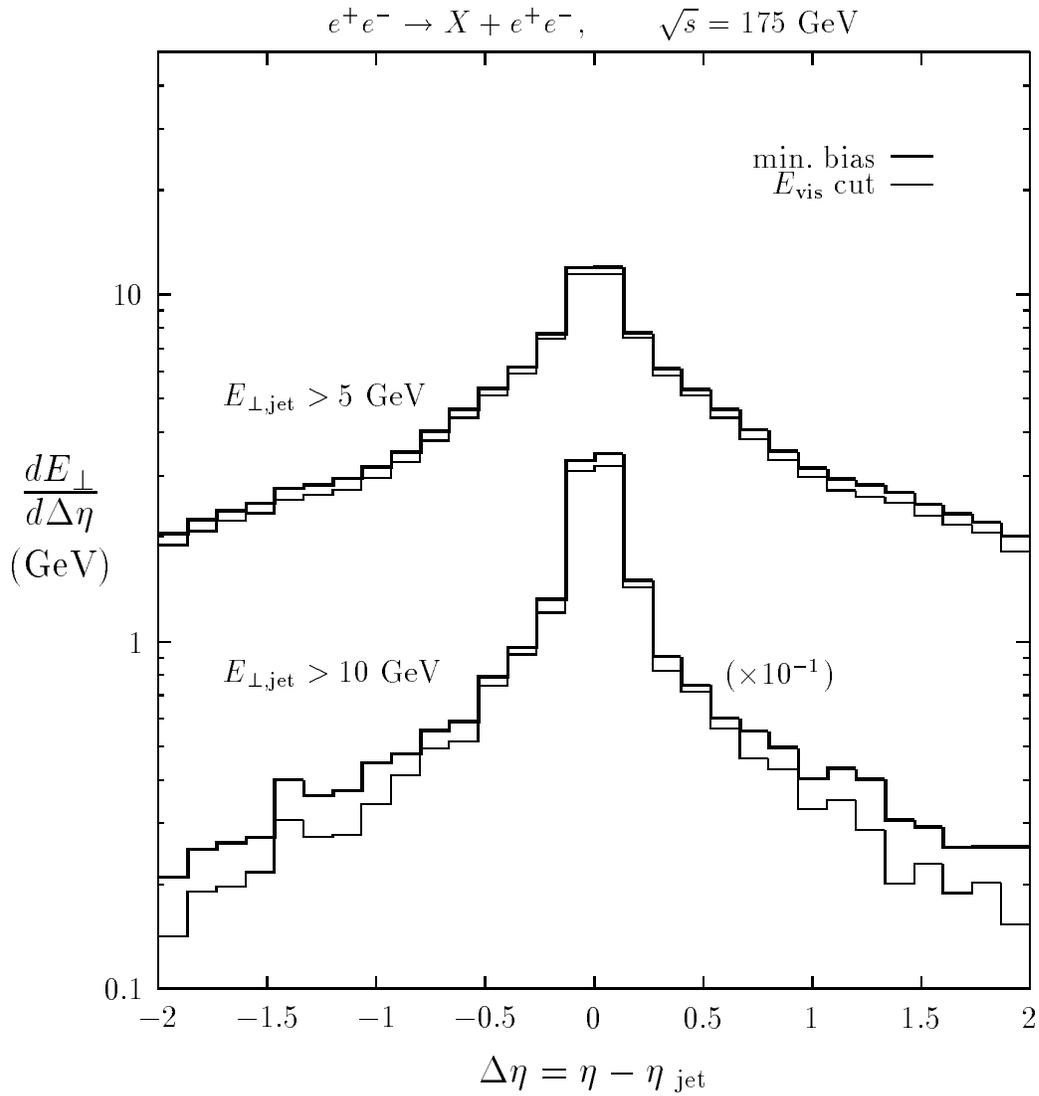

\caption{
Jet profiles in the pseudorapidity variable with and without the
visible energy cut described in the caption of
Fig.~\protect\ref{ggjetetv}.
\label{ggpro}
}
\end{figure}

To study the influence of an upper visible energy cut we plot in
Fig.~\ref{ggjetetv} the jet transverse energy cross section
together with the distribution obtained with the cut
$E_{\mbox{\scriptsize visible}} \leq$ 50 GeV within the acceptance region
$|\eta| \le 2.1$.
The 50 GeV cut serves to suppress jets coming from $Z^0$ decay.
For jets with transverse energies below 10 GeV this cut does not
change the $E_{\perp \mbox{\scriptsize jet}}$ distribution drastically.
The same is
found studying pseudorapidity jet profiles with and without the
visible energy cut in Fig.~\ref{ggpro}. Note that the jets become more
narrow with rising energy and also the pedestal under the jets
rises with rising jet transverse energy. This again is an effect
mainly due to the multiple soft and hard chains.

\begin{figure}
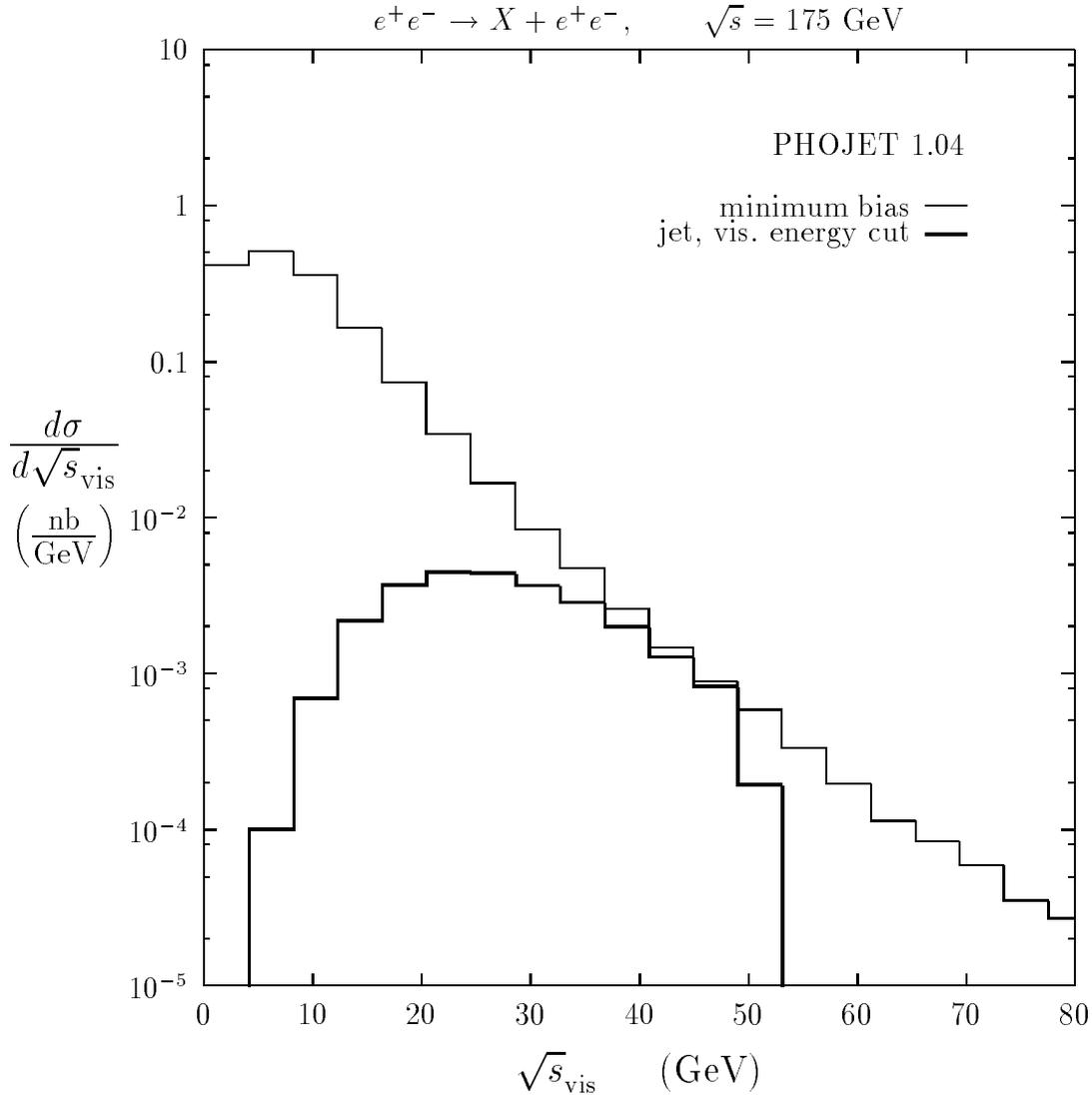

\caption{
Cross section as function of the visible energy
inside a calorimeter with $|\eta| \leq$ 2.1.
Only events  with ${\protect\sqrt s}_{\gamma\gamma} \geq$ 10 GeV are
sampled from the improved Weizs\"acker-Williams spectrum. In a
second curve we request jets with
\protect$E_{\perp \mbox{\scriptsize jet}} \geq$ 5 GeV and
impose a visible energy cut of 50 GeV.
\label{ggsvis}
}
\end{figure}

Finally, in Fig.~\ref{ggsvis} we plot the cross section as
function of the visible energy within the acceptance region.
Since we want to search jets with a
$E_{\perp \mbox{\scriptsize jet}} \geq$ 5 GeV cutoff,
we use a lower threshold for the photon-photon energy
${\sqrt s}_{\gamma\gamma}$ = 10 GeV. The plot gives the visible
energy distribution without restrictions and in a second curve
the  visible energy obtained if we demand at least one jet with
$E_{\perp \mbox{\scriptsize jet}} \geq$ 5 GeV
and impose the visible energy cut
$E_{\mbox{\scriptsize visible}} \leq$ 50 GeV
within the acceptance region given above.
We observe: nearly each event with a visible energy above 30
GeV contains at least one jet with
$E_{\perp \mbox{\scriptsize jet}} \geq$ 5 GeV.

%\clearpage

\section{Conclusions and summary}

The {\sc phojet} model can be used to calculate hadronic events in
hadron-hadron, photon-hadron and photon-photon collisions.
The model is found to agree  well with data in
hadron-hadron and photon-hadron collisions, the predictions
for photon-photon collisions do not need any new parameters.

Multiple soft and multiple hard interactions (minijets) lead to
a rise of the rapidity plateau, which agrees in hadron-hadron and
photon-hadron collisions very well with the rise of the
plateau observed experimentally.

Minimum bias hadron production in hadron-hadron,
photon-hadron and photon-photon collisions of the same CMS
energy is remarkably similar. To see this, one has to restrict
the comparison to inelastic events and to exclude also the
diffractively produced vector mesons in reactions involving
photons. The only striking differences appear in the transverse
momentum distribution or distributions, where the transverse
momentum behavior is essential. This difference can  be
understood to be due to the direct photon interaction
contribution and due to the photon structure function being
considerably harder than hadronic structure functions.

{\sc phojet} can be applied also to photon-photon
interactions in electron-hadron and electron-electron or
electron-positron colliders. Photon spectra according to the
equivalent photon approximation, according to beamstrahlung
spectra and backscattered laser radiation  in linear
electron-positron colliders are implemented at present. It is
easy to implement other similar photon spectra. With these
possibilities {\sc phojet} should be a tool suitable to study
the photon-photon background to other interesting reactions at
such colliders.

%\clearpage
\vskip 20mm

{\bf Acknowledgements}

The authors are grateful to P.~Aurenche for many helpful discussions
and suggestions.  They acknowledge the hospitality and support by LAPP,
Annecy-le-Vieux, most of the results presented were obtained
during stays at LAPP and they acknowledge the invitations to
sessions of LEP-II and linear collider study groups by the
conveners of these studies. One of the authors (R.E.) was supported by
the Deutsche Forschungsgemeinschaft under contract No. Schi
422/1-2.

%
%------- Bibliography ----------

%
%
%++++++++++++++++++++++++++++++++++++++++++++++++++++++++++++++++++++++
%

\clearpage
%
%
%\caption{
{\bf Table 1}
Comparison of average quantities characterizing hadron
production in nondiffractive
$p$--$p$, $\gamma$--$p$ and $\gamma$--$\gamma$
collisions at cms energies between 10 and 200 GeV. All average
transverse momenta are given in GeV/c.
%\label{pp200dpm}
\vskip 5mm
\begintable
$\sqrt s$ |Quantity      |$p$--$p$  | $\gamma$--$p$ |$\gamma$--$\gamma$ ~\cr
 10 | $n_{tot}$            |11.2   |11.1    |11.7 \cr
 10 | $n_{ch}$             |6.65   |6.53    |6.86 \cr
 10 | $n_{\pi^-}$          |2.17   |2.44    |2.88 \cr
 10 |$n_{\bar p}$          |0.027   |0.063    |0.11 \cr
 10 |$<p_{\perp  ch}>_{centr. \eta}$ |0.39   |0.38    |0.42 \cr
 10 |$<p_{\perp}>_{\pi^-}$ |0.32   |0.33    |0.36 \cr
 10 |$<p_{\perp}>_{\bar p}$|0.41   |0.43    |0.47 \cr
 10 |$<n>_{soft ch.} $     |1.16   |1.19    |1.22 \cr
 10 |$<n>_{minijets} $     |0.   |0.00004    |0.0020 \cr
 20 | $n_{tot}$            |16.4   |16.6    |17.1 \cr
 20 | $n_{ch}$             |9.64   |9.71    |10.00\cr
 20 | $n_{\pi^-}$          |3.44   |3.78    |4.18 \cr
 20 |$n_{\bar p}$          |0.086   |0.14    |0.20 \cr
 20 |$<p_{\perp  ch}>_{centr. \eta}$ |0.37   |0.38    |0.44 \cr
 20 |$<p_{\perp}>_{\pi^-}$ |0.32   |0.34    |0.38 \cr
 20 |$<p_{\perp}>_{\bar p}$|0.42   |0.45    |0.52 \cr
 20 |$<n>_{soft ch.} $     |1.26   |1.33    |1.31 \cr
 20 |$<n>_{minijets} $     |0.0003   |0.0025    |0.028 \cr
 50 | $n_{tot}$            |24.8   |26.5    |26.9 \cr
 50 | $n_{ch}$             |14.5   |15.5    |15.6 \cr
 50 | $n_{\pi^-}$          |5.49   |6.19    |6.53 \cr
 50 |$n_{\bar p}$          |0.21   |0.27    |0.34 \endtable
\clearpage

Continuation of Table 1
\vskip 5mm
\begintable
$\sqrt s$ |Quantity      |$p$--$p$  | $\gamma$--$p$ |$\gamma$--$\gamma$ ~\cr
 50 |$<p_{\perp  ch}>_{centr. \eta}$ |0.38   |0.40    |0.46 \cr
 50 |$<p_{\perp}>_{\pi^-}$ |0.33   |0.35    |0.40 \cr
 50 |$<p_{\perp}>_{\bar p}$|0.44   |0.47    |0.57 \cr
 50 |$<n>_{soft ch.} $     |1.50   |1.68    |1.44 \cr
 50 |$<n>_{minijets} $     |0.0096 |0.035   |0.17 \cr
 200 | $n_{tot}$            |40.1   |46.2    |47.5 \cr
 200 | $n_{ch}$             |23.3   |26.9    |27.6 \cr
 200 | $n_{\pi^-}$          |9.16   |10.94   |11.46\cr
 200 |$n_{\bar p}$          |0.46   |0.59    |0.67 \cr
 200|$<p_{\perp  ch}>_{centr. \eta}$ |0.40   |0.42    |0.48 \cr
 200 |$<p_{\perp}>_{\pi^-}$ |0.35   |0.38    |0.42 \cr
 200 |$<p_{\perp}>_{\bar p}$|0.47   |0.53    |0.64 \cr
 200 |$<n>_{soft ch.} $     |1.59   |1.87    |1.29 \cr
 200 |$<n>_{minijets} $     |0.17   |0.36    |1.01 \endtable
%1000 | $n_{tot}$            |73.8  |80.3    |78.3 \cr
%1000 | $n_{ch}$             |42.8   |46.6    |45.3 \cr
%1000 | $n_{\pi^-}$          |17.25   |19.12    |18.84 \cr
%1000 |$n_{\bar p}$          |0.99   |1.14    |1.17 \cr
%1000 |$<p_{\perp  ch}>_{centr. \eta}$ |0.46   |0.46    |0.49 \cr
%1000 |$<p_{\perp}>_{\pi^-}$ |0.39   |0.41    |0.44 \cr
%1000 |$<p_{\perp}>_{\bar p}$|0.56   |0.61    |0.67 \cr
%1000 |$<n>_{soft ch.} $     |1.20   |1.43    |0.92 \cr
%1000 |$<n>_{minijets} $     |1.33   |1.54    |2.16 \endtable
%\end{table}
%
 \clearpage
\vskip 5mm
%\caption{
{\bf Table 2}
 Average photon--photon energies and  weight factors in
$\mu b$ for the three photon spectra and two linear collider energies.
%\label{pp200dpm}
\vskip 5mm
\begintable
$\sqrt s$ |Photon spectrum |${\sqrt s}_{\gamma\gamma}$  |Weight ($\mu b$)~\cr
 500      | W.--W.         |50                          |0.0067 \cr
 500      | Beamst.        |16.7                        |0.038  \cr
 500      | B.Laser        |252                         |0.47   \cr
 1000     | W.--W.         |105                         |0.0075 \cr
 1000     | Beamst.        |41                          |0.026  \cr
 1000     | B.Laser        |509                         |0.59   \endtable
%\end{table}
%
%
%\clearpage
%
\vskip 5mm
%\caption{
{\bf Table 3}
Comparison of average quantities for LEP--II photon--photon
collisions with photon--photon collisions at
 ${\sqrt s}_{\gamma\gamma}$ = 25 GeV.
%\labEL{PP200DPm}
\vskip 5mm
\begintable
Quantity      |LEP--II $\gamma\gamma$  |$\gamma\gamma$ at 25 GeV ~\cr
 $n_{tot}$            |15.6   |17.2    \cr
 $n_{ch}$             |9.19   |10.1    \cr
 $n_{\pi^+}$          |3.74   |4.12    \cr
 $n_{\pi^-}$          |3.74   |4.12    \cr
$n_{\bar p}$          |0.17   |0.19    \cr
$<p_{\perp}>_{\pi^-}$ |0.37   |0.37    \cr
$<p_{\perp}>_{\bar p}$|0.50   |0.50    \cr
$<n>_{soft ch.} $     |1.36   |1.44    \cr
$<n>_{minijets} $     |0.035  |0.028   \cr
Weight ($\mu b$)      |0.0032 |   \endtable
%\end{table}
%
%
\clearpage
\clearpage
\begin{figure}
\caption{
We compare at the collision energy $\protect\sqrt s$ = 200 GeV the
radial $x_r = 2 E/\protect\sqrt s$ distribution in invariant form for all
charged hadrons produced in proton-proton, photon-proton and
photon-photon collisions. The calculation was done with
{\sc phojet} for inelastic collisions, excluding in
photon-proton collisions the $V$-$p$ and in photon-photon
collisions the $V$-$V$ diffractive contributions.
\label{phoxr200}
}
\end{figure}
\clearpage
\begin{figure}
\caption{
We compare at the collision energies $\protect\sqrt s$ = 20 GeV
and 200 GeV  in a log-plot the
 distribution  of the transverse energy $E_\protect{\perp}$ as function
 of the pseudorapidity $\eta$ for
  proton-proton, photon-proton and
photon-photon collisions. The calculation was done with
{\sc phojet} for inelastic collisions, excluding in
photon-proton collisions the $V$-$p$ and in photon-photon
collisions the $V$-$V$ diffractive contributions.
\label{phoetet}
}
\end{figure}
\begin{figure}
\caption{
We compare at the collision energies $\protect\sqrt s$ = 20 GeV
, 200 GeV and 1000 GeV in a linear plot the
 distribution  of the transverse energy $E_\protect{\perp}$ as function
 of the pseudorapidity $\eta$ for
  proton-proton, photon-proton and
photon-photon collisions. The calculation was done with
{\sc phojet} for inelastic collisions, excluding in
photon-proton collisions the $V$-$p$ and in photon-photon
collisions the $V$-$V$ diffractive contributions.
\label{phoetel}
}
\end{figure}
\clearpage

\end{document}